\documentclass[%
twocolumn,
 amsmath,amssymb,
aps,
prx,
]{revtex4-1}

\usepackage{graphicx,epsfig,epstopdf}
\usepackage{amsmath}
\usepackage{color}
\usepackage{caption}
\usepackage{subcaption}

\usepackage{tabularx}

\def\e{\begin{equation}}
\def\f{\end{equation}}
\def\_#1{{\bf #1}}
\def\.{\cdot}

\begin{document}

\title{\textcolor{black} {Parallel Optical Computing Based on  MIMO Metasurface Processors with Asymmetric Optical Response } } 

\author{Amirhossein Babaee$^1$, Ali Momeni$^1$, Ali Abdolali$^1$, Romain Fleury$^2$}
 \affiliation{%
 	$^1$Applied Electromagnetic Laboratory, School of Electrical Engineering, Iran University of Science and Technology, Tehran, Iran\\
 	$^2$Laboratory of Wave Engineering, Swiss Federal Institute of Technology in Lausanne (EPFL),\\ CH-1015 Lausanne, Switzerland\\
 }
\begin{abstract}
We present a polarization-insensitive  metasurface processor to perform spatial asymmetric filtering of an incident optical beam, thereby allowing for real-time parallel optical processing. To enable massive parallel processing, we introduce a novel Multi Input-Multi Output (MIMO) computational metasurface with asymmetric optical response that can perform spatial differentiation on two distinct input signals regardless of their polarization. In our scenario, two distinct signals set in x and y directions, {parallel and perpendicular to the incident plane,} illuminate simultaneously the metasurface processor, and the resulting differentiated signals are separated from each other via appropriate Spatial Low Pass Filters (SLPF).
By leveraging Generalized Sheet Transition Conditions (GSTCs) and surface susceptibility tensors, {we design} an asymmetric meta-atom augmented with normal susceptibilities \textcolor{black}{to reach asymmetric optical response at normal beam illumination}
. Proof-of-principle simulations are also reported along with the successful realization of signal processing
functions.  The proposed metasurface overcomes major shortcomings imposed by previous studies such as large architectures arising from the need of additional subblocks, slow responses, and most importantly, supporting only a single input with a given polarization.
Our results set the path for future developments of material-based analog computing using efficient and easy-to-fabricate MIMO processors for compact, fast, and integrable computing elements without any Fourier lens.
\end{abstract}

\maketitle

\section{Introduction}
With the expeditious development of technology in today's communications systems, signal and image processing has gained a lot of attention in the past decade \cite{white,goodman2005introduction,stark1982applications}. The idea of signal processing by analog computing has been known for a long time, but in the late 20th century, when the digital revolution began, digital computation sat in place of analog computation. Although it was a huge breakthrough, these digital systems suffer from serious restrictions such as data conversion loss and operational speed \cite{doi:10.1080/00043249.1990.10792698}. Analog solutions therefore emerged again and proved to be advantageous for specific tasks, for example the processing of large size images \cite{Silva160,2015NaPho...9..704S}. Optical signal processing, in particular, largely overcomes the serious
limitations of digital systems in term of speed and power consumption \cite{2015NaPho...9..704S}.
Therefore, designing fast, integrated, and ultra-low power consumption optical devices with high-throughput  is one of the key necessities to develop today's modern optical analog processing, and move towards commercial applications with real-time performance.

Spatial optical analog computing based on computational metamaterials is generally based on two major methods: Green’s Function (GF) method and metamaterial surfaces (metasurface), with two additional sub-blocks to apply Fourier and Inverse Fourier Transforms \cite{abdollahramezani2017dielectric,Babashah:17,pors2015analog}. Both approaches, however, suffer from the large final size of the system  \cite{Silva160,Babashah:17, abdollahramezani2017dielectric}. Over the past few years, in order to sidestep the drawbacks associated with the Fourier transform sub-blocks, 
optical analog computing based on a single metasurface has attracted particular attention
 as an artificial real-time and high-throughput thin film computer. This allowed migrating from free-space and bulky systems into conceptually subwavelength-sized meta-atoms to perform mathematical operations \cite{Youssefi:16,PhysRevLett.121.173004,ZANGENEHNEJAD2018338,PhysRevApplied.11.064042,Zangeneh_Nejad_2019,Abdolali:273146,PhysRevApplied.11.034043,Zhu_2017,ZHOU2020124674,doi:10.1021/acs.nanolett.9b02477,doi:10.1002/adom.201901523,ss,PhysRevLett.123.013901,Guo:18,PMID:31391468,Karimi:20,Zhou11137,rajabalipanah2020spacetime}. Metasurfaces that efficiently manipulate the optical wave in the spatial
domain are synthesized via several approaches such as GSTCs and susceptibility tensors, which provide an insightful vision for the engineering of meta-atoms with specific angular scattering properties  \cite{PhysRevApplied.11.064042,Abdolali:273146,Yu2014FlatOW,Yu333,doi:10.1021/nl3032668,Kildishev1232009,8852838}. Due to their benefits in terms of low-profile, low sensitivity to absorption losses, and ease of fabrication, metasurfaces are interesting practical alternatives for bulky solutions, especially when it comes to applications including tunable/broadband  scattering manipulation \cite{momeni2019tunable,ROUHI2019125,Shahid2020,Kiani:20,doi:10.1002/andp.201700310,doi:10.1021/acsomega.9b02195,Hosseininejad2019ReprogrammableGM,KARGAR2020125331,article1} or antenna engineering  \cite{Hosseininejad2019DigitalMB,PhysRevApplied.11.044006,8731665,article}. 
\begin{figure*}[t]
	\centering{
		\label{fig:3}%
		\includegraphics[scale=.6]{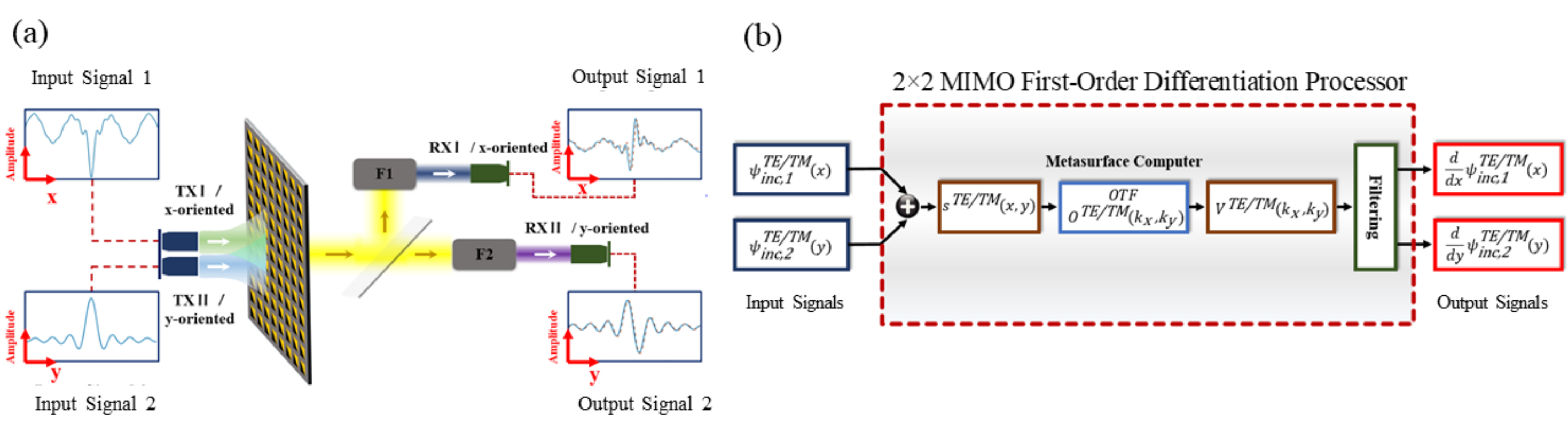}}%
	\caption{ \textcolor{black}{(a) Schematic sketch of the proposed spatial MIMO metasurface 
		processor for performing real-time parallel optical signal processing. (b) The block diagram view of the $2\times2$ MIMO first-order differentiation processor. Input signal 1, $\psi^{\text{TE/TM}}_{\text{inc,1}}(x)$, and Input signal 2, $\psi^{\text{TE/TM}}_{\text{inc,2}}(y)$, have x- and y-variations, respectively. The input signals illuminate normally and simultaneously the metasurface processor and the transmitted waves are separated by F1 and F2 which are spatial low pass filters related to the $k_x$ and $k_y$ directions, respectively. Finally, $\frac{d}{dx}\psi^{\text{TE/TM}}_{\text{inc,1}}(x)$ and $\frac{d}{dy}\psi^{\text{TE/TM}}_{\text{inc,2}}(y)$  as output signals are obtained. {The processor works equally for TE or TM polarisation.} }}
\end{figure*}

In order to be relevant to the widest possible class of signal processing operations, 
it is momentous that designed optical systems can operate filtering operations on multiple inputs, especially to develop massive parallel-processing schemes. Recently, some proposals have theoretically introduced parallel signal processing for inputs at different angles or two orthogonal polarizations \cite{PhysRevApplied.11.064042,Abdolali:273146,PhysRevApplied.11.034043}.
Although numerous efforts have been made to expand the functionalities of wave-based signal processing systems, a solution allowing for parallel signal processing without the stringent requirements of working at different incidence angles or given polarization states has not yet been proposed.
In addition, another resurgent challenge is creating a metasurface processor with an asymmetric response with respect to transverse momentum, namely one that distinguishes between $-k_t$ and $k_t$ components of normally-incident beams. Such odd Optical Transfer Functions (OTFs) are needed for first order differentiation and edge detection \cite{4767851,doi:10.1098/rspb.1980.0020}. Indeed, among all mathematical operations, spatial
differentiation is a fundamental mathematical operation used in many fields of science or engineering, and it is appealing for real-time image processing such as
image sharpening and edge-based
segmentation, with broad applications ranging from microscopy and medical imaging to industrial
inspection and object detection \cite{doi:10.1098/rspb.1980.0020} . 

In this paper, we design a Multiple-Inputs Multiple-Outputs (MIMO) computational metasurface for performing parallel optical processing with odd OTF on the  transmission operational mode at normal beam illumination. We propose and demonstrate a real time parallel polarization-insensitive  metasurface computer which can act as  first-order differentiation operator for both TE and TM states. Systematically speaking, here, we propose 2$\times$2 MIMO first-order differentiation processor for both orthogonal polarizations. The inputs and outputs correspond to different beam amplitude variations along orthogonal directions for a given polarization and angle.
Using GSTCs and susceptibility tensors, we show that the presence of normal susceptibility components is required for breaking the even transmission symmetry at normal illumination. Additionally, based on relationships between the structural symmetries of the meta-atoms and the corresponding symmetries of their angular scattering response, we propose a simple meta-atom with geometrical symmetry with respect to the x=y line, that enables first-order derivation for both input signals and both orthogonal polarizations. The  waves transmitted through the metasurface processor are then processed by simple Spatial Low Pass Filters (SLPFs) \textcolor{black}{ {in order to extract and separate} the  output signal with x-variations from {the one} with y-variations.}  In order to demonstrate the concept, we {provide} a numerical example based on arbitrary input signals. The synthesized metasurface paves the path towards the implementation of MIMO spatial analog mathematical systems to accelerate optical signal and image processing routines.

\section{GSTC and Meta-atom Design}

The general concept of wave-based MIMO signal processing for normally incident beams {is} summarized in \textcolor{blue}{Fig. 1}. \textcolor{black}{{We c}onsider the case of {a} $2\times2$ MIMO first-order differentiation processor{. T}wo distinct input signals, $\psi^{\text{TE/TM}}_{\text{inc,1}}(x)$ and $\psi^{\text{TE/TM}}_{\text{inc,2}}(y)$, with x- and y-variations, illuminate simultaneously the metasurface processor for { a given (TE or TM) polarization}. By properly design{ing} the metasurface processor, the transmitted waves {possess the same polarization as the input and}  contain {a sum of their} first-order differentiated signals. We {then utilize} two SLPF, along the $k_x$ and $k_y$, to extract the desired output signals{, namely} $\frac{d}{dx}\psi^{\text{TE/TM}}_{\text{inc,1}}(x)$ and $\frac{d}{dy}\psi^{\text{TE/TM}}_{\text{inc,2}}(y)$.} Theoretically speaking, we start from a reciprocal passive metasurface processor consisting of a periodic and homogeneous array of polarizable meta-atoms in the $z=0$ plane, which acts as a two-dimensional electromagnetic discontinuity on the incident field created by external sources (see \textcolor{blue}{Figs. 2}). Throughout the paper, we will assume a time-harmonic dependence of the form $e^{j\omega t}$, where $j^2=-1$ is the imaginary unit.

A metasurface, described by a set of surface susceptibility components $({{\overline{\overline{\chi }}}_{\text{ee}}},{{\overline{\overline{\chi }}}_{\text{em}}},{{\overline{\overline{\chi }}}_{\text{me}}},{{\overline{\overline{\chi }}}_{\text{mm}}})$, can generate output fields with the desired transverse spatial dependency of \textcolor{black}{$\psi _{\text{ref/tran}}^{\text{TE/TM}}({{x,y}})$} in transmission/reflection mode when arbitrary input fields having the spatial dependence \textcolor{black}{$\psi _{\text{inc}}^{\text{TE/TM}}({{x,y}})$} excite it. In fact, \textcolor{black}{$\psi _{\text{inc}}^{\text{TE/TM}}({{x,y}})$} and \textcolor{black}{$\psi _{\text{ref/tran}}^{\text{TE/TM}}({{x,y}})$} can be consider{ed} as the input and output signals of our linear system, where, the angular EM response of the metasurface processor determines the corresponding OTF, \textcolor{black}{$O(k_x,k_y)$}, in the spatial Fourier domain. Actually, the OTF represents the response to plane waves at different angles of incidence, providing a
useful representation of the properties of optical metasurfaces \cite{PhysRevLett.123.013901} .

\begin{figure}[t!]%
	\centering
	{%
		\label{fig:3}%
		\includegraphics[scale=.3]{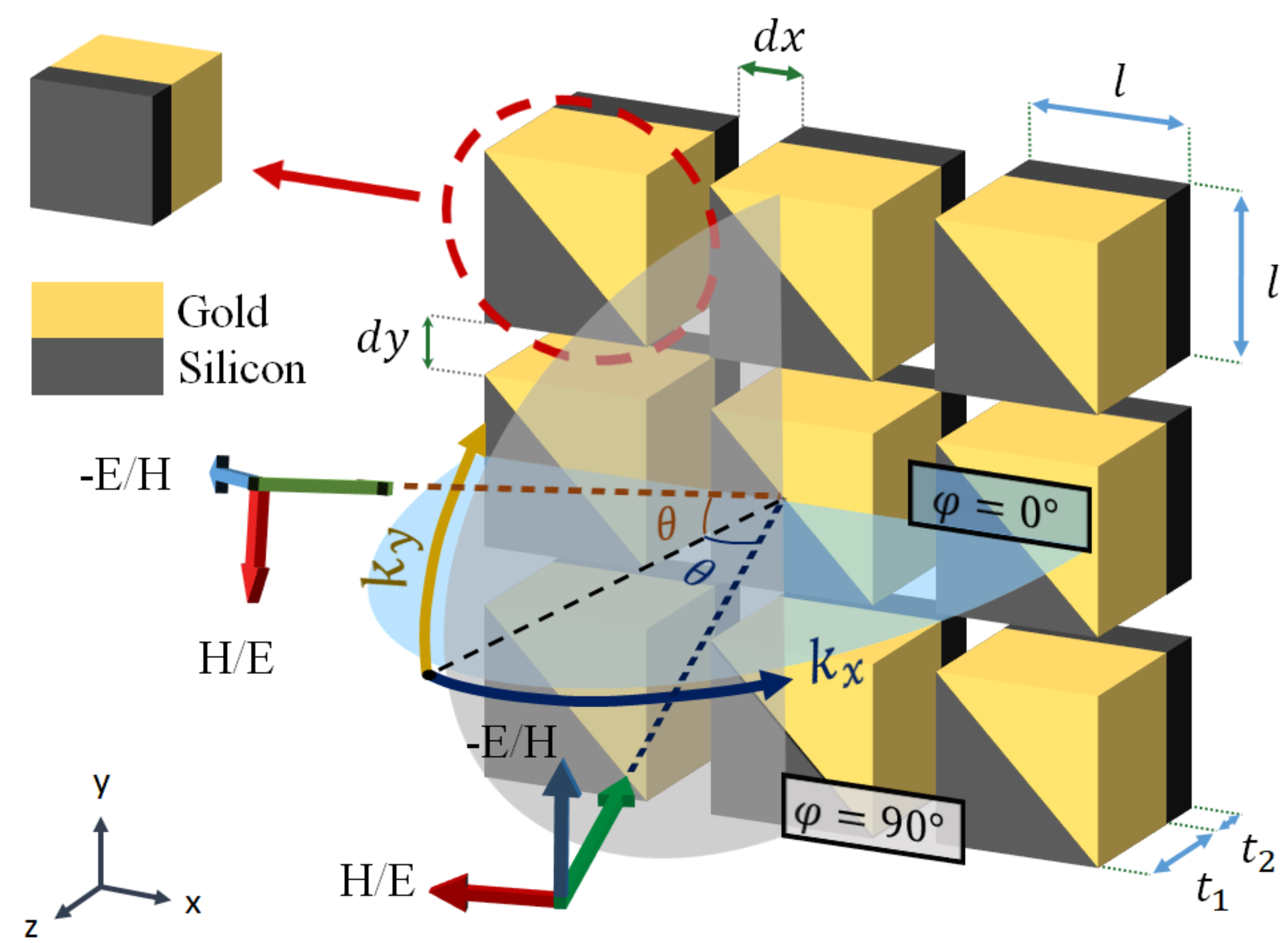}}%
	\caption{  Realization of the MIMO metasurface computer at an optical wavelength $\lambda_0=48.935 \mu m$. The schematic shows an array of designed meta-atoms in the x-y plane. The metasurface is excited by TM or TE-polarized electromagnetic waves.
		The	geometrical parameters are $dx=dy=4 \mu m$, $l=15 \mu m$, $t_1=11 \mu m$ and $t_2=4 \mu m$ .   }
	\label{fig:5}
\end{figure}

Here, the solutions can be computed by \textcolor{black}{${{\psi}_{\text{ref/tran}}}\left( x,y \right)={{F}^{-1}}\left[ \tilde{O}\left( {{k}_{x},k_y} \right)\times \\
F\left( {{\psi}_{\text{inc}}}\left( x,y \right) \right) \right]$} in which $F$ and $F^{-1}$ represent the operation of Fourier and inverse Fourier transform, respectively, and $k_{x}$ and $k_y$ denote the spatial frequency variable in the Fourier space. 
Commonly, the OTF provided by an array of polarizable meta-atoms has a tensorial format for both  orthogonal polarizations which can be written as  
\textcolor{black}{\begin{equation}
\overline{\overline{{{O}}}}^{\text{TE/TM}}({{k}_{x},k_y})\equiv \left[ \begin{matrix}
{{{\tilde{u}}}^{\text{TE/TE}}}\left( {{k}_{x},k_y} \right)\, & {{{\tilde{u}}}^{\text{TE/TM}}}\left( {{k}_{x},k_y} \right)\,\,  \\
{{{\tilde{u}}}^{\text{TM/TE}}}\left( {{k}_{x},k_y} \right)\,\, & {{{\tilde{u}}}^{\text{TM/TM}}}\left( {{k}_{x},k_y} \right)\,  \\
\end{matrix} \right]\,\,\,
\end{equation}}

where \textcolor{black}{$\tilde{u}(k_x,k_y)$} refers to the functionality of the reflection (R) or transmission (T) coefficient of the metasurface processor from the incident wave angle. The first and second superscripts also represent the polarization of the input and output waves, respectively. 
The spatial OTF belonging to the metasurface computer can be extracted using the GSTCs formalism in which the metasurface transition conditions read \cite{8852838,doi:10.1063/1.4972195,art3icle,refId0,8259235}
\begin{align}
& \hat{z}\times \Delta \textbf{H}=j\omega {{\textbf{P}_{\parallel }}}-\hat{z}\times {{\nabla }_{\parallel }}{{\textbf{M}}_{z}} \\ 
& \Delta \textbf{E}\times \hat{z}=j\omega \mu {{\textbf{M}_{\parallel }}}-{{\nabla }_{\parallel }}\left(\frac{{{\textbf{P}}_{z}}}{\varepsilon_0 }\right)\times \hat{z}
\end{align}
in which, $\Delta \textbf{E}$ and $\Delta \textbf{H}$ are the difference of the electric and magnetic fields on both sides of the metasurface, respectively. $\textbf{P}$ and $\textbf{M}$ represent the electric and magnetic polarization densities induced on the metasurface. The susceptibility tensor components relate the polarization densities to  the average electric and magnetic fields on both sides of the metasurface processor as  
$\textbf{P}=\varepsilon_0 {{{\bar{\bar{\chi }}}}_{\text{ee}}}{{\textbf{E}}_{\text{av}}}+{{{\bar{\bar{\chi }}}}_{\text{em}}}\sqrt{\mu_0 \varepsilon_0 }{{\textbf{H}}_{\text{av}}}$ and $ 
\textbf{M}={{{\bar{\bar{\chi }}}}_{\text{mm}}}{{\textbf{H}}_{\text{av}}}+{{{\bar{\bar{\chi }}}}_{\text{me}}}\sqrt{\frac{\varepsilon_0 }{\mu_0 }}{{\textbf{E}}_{\text{av}}},  $ ~~~Here, $\epsilon_0$ and $\eta_0$ are the permittivity and the characteristic impedance of free-space, respectively. 
In the most general case, each susceptibility tensor appearing in Equations. (2) and (3) include both longitudinal and tangential components, i.e. 36 susceptibilities.
\begin{figure}[t!]%
	\centering
	{%
		\label{fig:3}%
		\includegraphics[scale=.3]{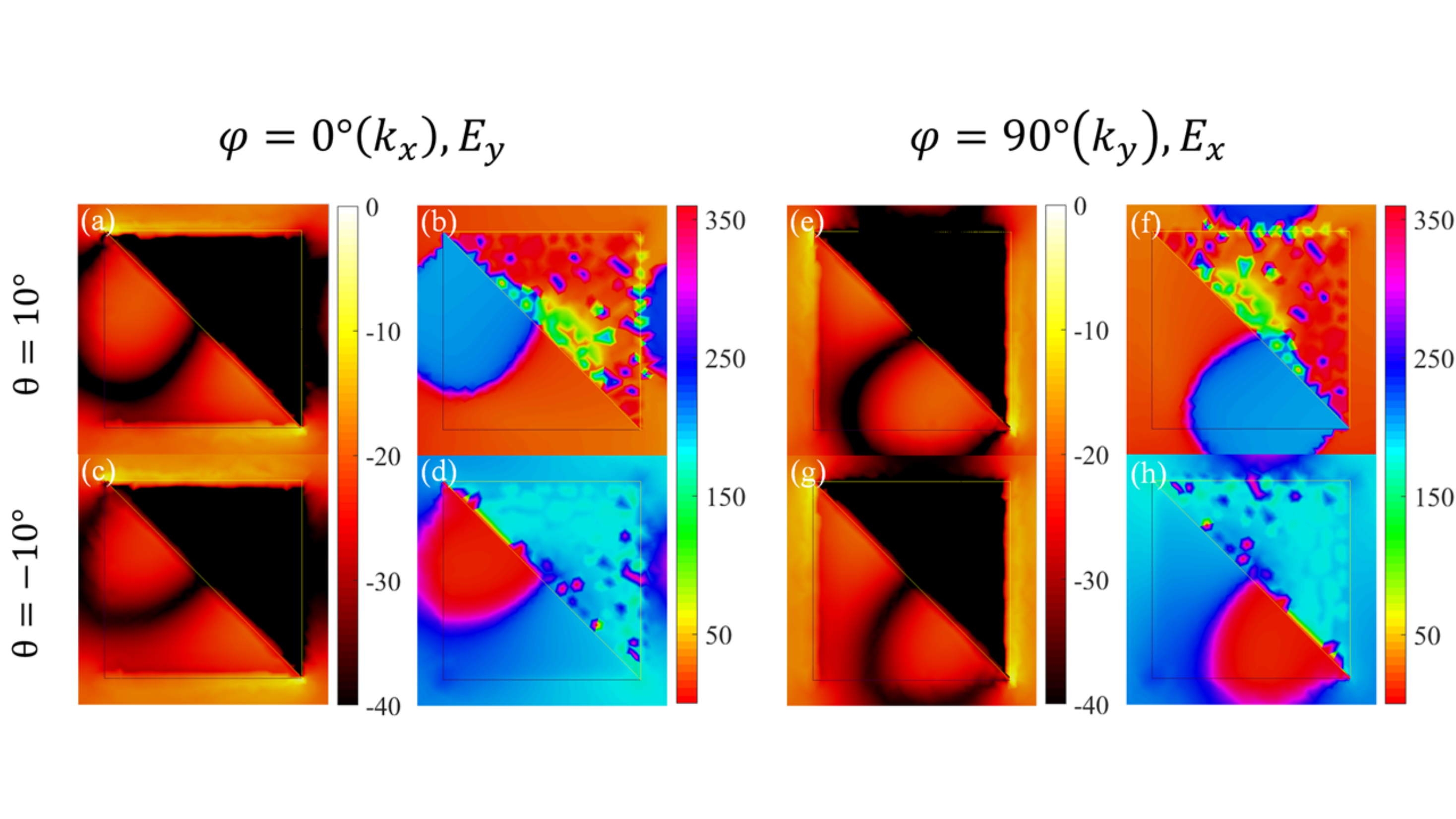}}%
	\caption{ Electric field profiles at \textcolor{black}{plane z = 0 {(}top surface of meta-atoms{)}} when $\theta = 10^\circ$ and  $-10^\circ$. (a), (c), (e) and (g) are amplitude of electric fields for $\theta = 10^\circ$ and  $-10^\circ$. Similarly, other plots contain phase information of electric fields.}
	\label{fig:5}
\end{figure}
\begin{figure*}[t]%
	\centering
	{%
		\label{fig:3}%
		\includegraphics[scale=.53]{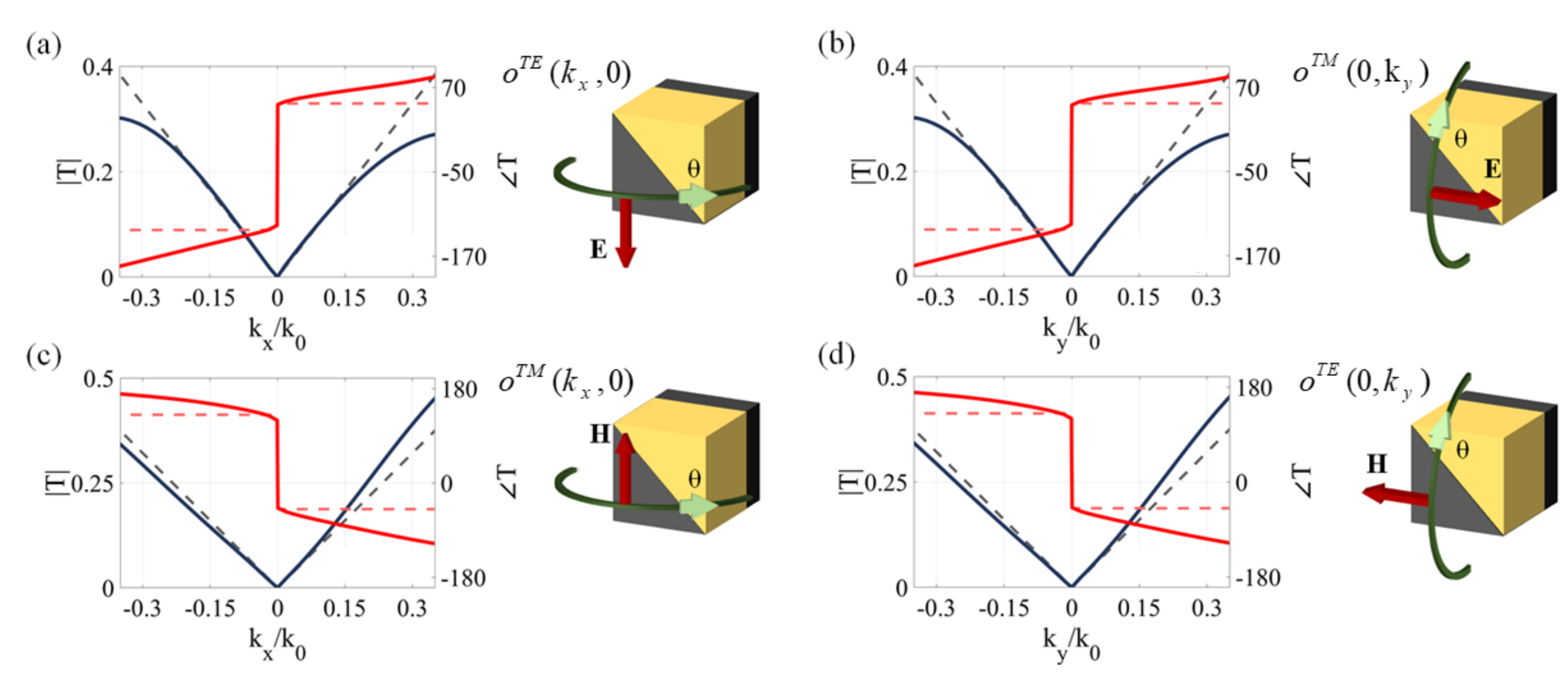}}%
	\caption{Synthesized OTF. (a) and (c) The amplitude (black) and phase (red) of the synthesized OTF$(k_x,0)$ associated with the first-order differentiation realized by the metasurface processor for TE and TM polarization, respectively. Also,
		(b) and (d) are similar plots for OTF$(0,k_y)$. The synthesized and ideal OTFs are indicated with solid and dashed lines, respectively.}
	\label{fig:5}
\end{figure*}
As we see in Equations. (2) and (3) if we consider a homogeneous metasurface without normal polarization  ($\text{P}_z=\text{M}_z=0$), only purely tangential polarizations exist, and no asymmetric function with respect to x-coordinate ($k_x$ for example ) is allowed. In fact, in this circumstance, the synthesized OTF has an even response (${k_x}^2$ for example) respect to x-coordinate. However, when we involve the $\text{P}_z$ and $\text{M}_z$ terms, according to Equations. (2) and (3), the $\partial_x=jk_x$ (or $\partial_y=jk_y$) operators in ${{\nabla }_{\parallel }}(.)$ make a spatial asymmetric OTF possible. To delve more into it and have an better awareness of how the angular scattering response of a metasurface depends on its normal susceptibilities, we begin the study with a simplified but pertinent scenario. \textcolor{black}{{W}e synthesize the metasurface processor {with an} expected performance {described by an} OTF {specified on} $(k_x,0)$ and $(0,k_y)$. For brevity and avoiding the complexity of relations, we will write the relations, Reflection and Transmission of metasurface, only for $(k_x,0)$. {Naturally}, a similar procedure can be mimicked for $(0,k_y)$. } Without loss of generality, we focus on TM-polarized illumination. Keeping the polarization preservation criterion in mind, the susceptibility components that may be excited in our case are 
\begin{align}
& \overline{\overline{{{\chi }}}}_{\text{ee}}=\left( \begin{matrix}
\begin{matrix}
\begin{matrix}
\chi _{\text{ee}}^{xx}  \\
0  \\
\chi _{\text{ee}}^{zx}  \\
\end{matrix} & \begin{matrix}
0  \\
0  \\
0  \\
\end{matrix}  \\
\end{matrix} & \begin{matrix}
\chi _{\text{ee}}^{xz}  \\
0  \\
\chi _{\text{ee}}^{zz}  \\
\end{matrix}  \\
\end{matrix} \right), \,\,\text{  }\overline{\overline{{{\chi }}}}_{\text{mm}}=\left( \begin{matrix}
\begin{matrix}
\begin{matrix}
0  \\
0  \\
0  \\
\end{matrix} & \begin{matrix}
0  \\
\chi _{\text{mm}}^{yy}  \\
0  \\
\end{matrix}  \\
\end{matrix} & \begin{matrix}
0  \\
0  \\
0  \\
\end{matrix}  
\end{matrix} \right)
\end{align}
~~The reciprocity conditions enforce $\bar{\bar{\chi }}_{\text{ee}}^{T}={{\bar{\bar{\chi }}}_\text{{ee}}},\,\,\,\bar{\bar{\chi }}_{\text{mm}}^{T}={{\bar{\bar{\chi }}}_{\text{mm}}}$. In this case, the reflection and transmission coefficients can be expressed in terms of the metasurface susceptibilities as shown in \cite{8852838}, and are given by \textcolor{black}{
\begin{align}
& \gamma =2[k_{z}^{2}\chi _{\text{ee}}^{xx}+k_{x}^{2}\chi _{\text{ee}}^{zz}+{{k_0}^{2}}\chi _{\text{mm}}^{yy}]+{{k_0}^{2}}(\chi _{ee}^{xx}\chi _{\text{mm}}^{yy}) 
-\\\nonumber&j{{k}_{z}}[k_{x}^{2}({\chi _{\text{ee}}^{xz}}^2-\chi _{\text{ee}}^{xx}\chi _{\text{ee}}^{zz})+4]. \\ 
& {\textbf{T}}({{k}_{x},0})=\frac{j{{k}_{z}}}{\gamma }\{k_{x}^{2}({\chi _{\text{ee}}^{xz}}^2-\chi _{\text{ee}}^{xx}\chi _{\text{ee}}^{zz})-4 
+4j{{k}_{x}}\chi _{\text{ee}}^{xz}\\\nonumber
&-{{k_0}^{2}}\chi _{\text{ee}}^{yy}\chi _{\text{mm}}^{xx}\}. \\\nonumber 
& \textbf{{R}}({{k}_{x},0})=\frac{2}{\gamma}\{k_{x}^{2}\chi _{\text{ee}}^{zz}-k_{z}^{2}\chi _{\text{ee}}^{xx} +{{k_0}^{2}}\chi _{\text{mm}}^{yy}\}. \\ 
\end{align}}
~~~As can be noticed from Equations. (5) and (6), the presence of non-zero $\chi _{\text{ee}}^{xz}$ component makes the transmission transfer function odd with respect to the $k_x$ variable, creating an asymmetric function of $\theta$. The other normal susceptibility components cannot be use to break the angular symmetry of the transmission transfer function, e.g. $\chi _{\text{ee}}^{zz}$ induces a term proportional to $k_x^2$ in the relations, which implies an even-symmetric function of $\theta$. We conclude that the normal polarizability component  $\chi _{\text{ee}}^{xz}$ is mandatory for breaking the  angular symmetry of the transmission of the metasurface.

From a realization point of view, there are fruitful relationships between the structural symmetries of the metasurface scattering meta-atoms and the corresponding symmetries of their angular scattering response. Geometrically, three types of symmetries are conceivable for the constituent meta-atoms: a reflection symmetry through the z-axis ($\sigma_{z}$), a $180^{\circ}$-rotation symmetry around the y-axis ($C_2$), and a reflection symmetry through the x-axis ($\sigma_{x}$) \cite{8852838}. Regarding a reciprocal metasurface, the angular spectrum of the reflection coefficient exposes a $\sigma_{z}$ symmetry , while that of the  transmission coefficient has a $C_2$ symmetry. Most importantly, a metasurface with both normal and tangential susceptibilities are macroscopically achieved when the occupying meta-atoms do not microscopically render any geometrical symmetry.

In fact, when the meta-atoms are deprived of mentioned geometrical symmetry the resultant metasurface  will present $\chi_{ee}^{xz}$ (or $\chi_{ee}^{zx}$).  Breaking both vertical and horizontal mirror symmetries of meta-atom is therefore required for realizing the asymmetric OTF response \cite{8852838}. In this line of thought, we design an asymmetric meta-atom  \textcolor{blue}{Fig. 2b} comprising of two materials (gold \cite{hand} and silicon ($\epsilon_r=12$)), for operation at $\lambda_0=48.935 \mu m$. The presented meta-atom  would neither be $\sigma_{z}$ symmetric nor $C_2$ symmetric. In fact, we break both vertical and horizontal mirror symmetries to enable a non-zero normal polarization, $\chi _{\text{ee}}^{xz}$, reaching an asymmetric OTF.

Another important point about the structural symmetry of the designed meta-atom is that the meta-atom is symmetric respect to x=y line. This symmetrical feature is of importance for reaching parallel optical signal computing. This point can be figured out from \textcolor{blue}{Fig. 3} which shows the amplitude and phase of electric field profiles at plane z=0 for both $\theta=10^\circ$ and $-10^\circ$. By comparing the \textcolor{blue}{Fig. 3 a and b} with \textcolor{blue}{Fig. 3 e and f}, the symmetrical feature respect to x=y line is obvious. In fact, as we expected, the mirror image of \textcolor{blue}{Fig. 3 a and b} can be seen in \textcolor{blue}{Fig. 3 e and f}. A similar discussion for \textcolor{blue}{Fig. 3 c and d} and \textcolor{blue}{Fig. 3 g and h} is valid.
In summary, due to this geometrical symmetry, we expect effectively to attain the same transmission  response in x and y directions which \textcolor{black}{enables} parallel optical signal processing. 

In the following, we will demonstrate some possible wave-based functionalities that can
be unlocked by the proposed metasurface processor, as the fundamental block realizing the operator of choice. One of the most important mathematical functions that	has been less explored in the literature is the first-order differentiation operator whose spatial OTF can be written as $O(k_x,0) = \alpha jk_x$ and $O(0,k_y) =\beta jk_y$, \textcolor{black}{ where $\alpha$ and $\beta$ are complex constant coefficients which represent the gain values of the first-order differentiation operators.}

\textcolor{blue}{Fig. 4} displays the synthesized and required OTF for transmitting the first-order derivation of the	input field-profile for both TE and TM polarizations The structural specifications of the meta-atom leads to resonance for normal incident waves ($k_x=k_0$).  \textcolor{black}{As we can see in \textcolor{blue}{Fig. 4}, the amplitude of the transmission coefficient is infinitesimal at the  normal incident angle and it linearly increases when the incident angle changes. The 180 phase difference degrees between ${{k_t}>0}$ and ${{k_t}<0}$ of OTF indicates asymmetric, odd, angular response of OTF .} The results are simulated using CST full-wave commercial software. 

Here, \textcolor{black}{ as we discussed above}, the geometry of the proposed meta-atom enables us to have a same synthesized OTF for y directions \textcolor{black}{ which is important for our goal (see \textcolor{blue}{Fig. 4}).}
The proposed meta-atom can elaborately mimic the required $k_{x}$ and $k_{y}$-dependency of  first-order differentiator (see \textcolor{blue}{Fig. 4}). 
The phase and amplitude of the resulting transmission coefficient are plotted for both orthogonal polarizations in \textcolor{blue}{Figs. 4 a-d}. An excellent agreement between the synthesized OTF and the exact OTF has been achieved as long as the normalized spectral beamwidth of the input signals lies within $|W/k_0|< 0.35$. \textcolor{black}{As we can see in \textcolor{blue}{Fig.4}, the values of $\alpha$ and $\beta$ are
$|\alpha|\approx|\beta| \approx 1$}

Also, it is clear from electric field profile point of view which is plotted in \textcolor{blue}{Fig. 3}; by comparing the \textcolor{blue}{Fig. 3 a and b} with \textcolor{blue}{Fig. 3 c and d}, as we expect, we see the same amplitude and 180 phase difference between the $\theta=10^\circ$ and $\theta=-10^\circ$.  The same discussion for \textcolor{blue}{Fig. 3 e and f} and \textcolor{blue}{Fig. 3 g and h} for $\phi=90$ is valid.

\section{Real-time Parallel Optical Signal Processing   }

\begin{figure}[t]%
	\centering
	{%
		\label{fig:3}%
		\includegraphics[scale=.45]{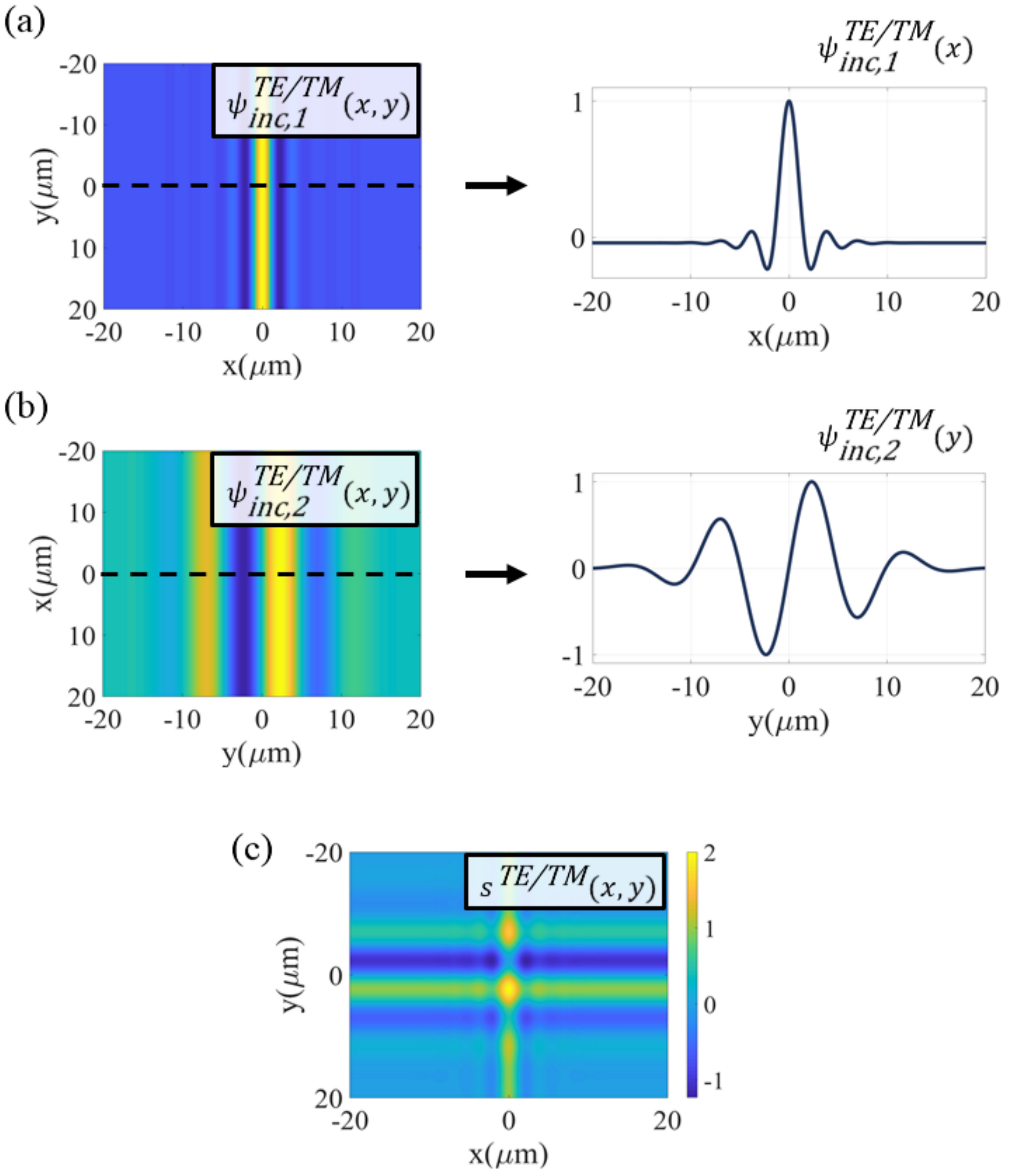}}%
	\caption{Input signals.  (a) and (b) The first and second arbitrary incident field profiles, respectively. (c) Combination of both input signals in x-y plane for both orthogonal polarization states. }
	\label{fig:5}
\end{figure}
In this section, we
explore a 2$\times$2 MIMO first-order differentiation processor, which enables independent parallel
channels for signal processing, a key result of this paper.
\begin{figure*}[t]%
	\centering
	{%
		\label{fig:3}%
		\includegraphics[scale=.6]{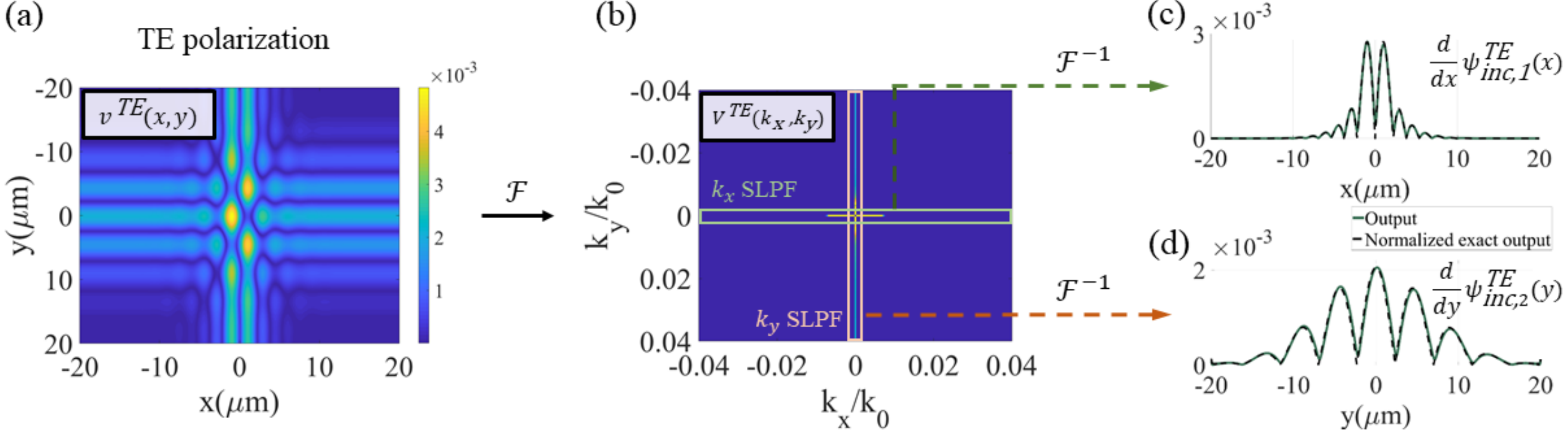}}%
	\caption{Output signals. The TE-polarized transmitted image in space domain (a) and in Fourier domain (b). (c) and (d) are  extracted output signals after passing from $k_x$-SLPF and $k_y$-SLPF, respectively.    }
	\label{fig:5}
\end{figure*}

Our final goal is performing the first-order differentiation simultaneously on two distinct input signals, $\psi_{\text{inc,1}}^{\text{TE/TM}}(x)$ and $\psi_{\text{inc,2}}^{\text{TE/TM}}(y)$, regardless of the polarisation chosen {(TE/TM means that we have the freedom to choose either TE or TM polarisation)}. In fact, we start from a combination of two arbitrary input signals (${s^{\text{TE/TM}}(x,y)}=\psi_{\text{inc,1}}^{\text{TE/TM}}(x)+\psi_{\text{inc,2}}^{\text{TE/TM}}(y) $) created by two sources, in which one of them has an x variation($\psi_{\text{inc,1}}^{\text{TE/TM}}(x)$) and another one has a y variation ($\psi_{\text{inc,2}}^{\text{TE/TM}}(y)$). Therefore, by using a Fourier Transform, we can theoretically  write  
\begin{align}
S^{\text{TE/TM}}(k_x,k_y)=\Psi_{\text{inc,1}}^{\text{TE/TM}}(k_x)\delta(k_y)+ \Psi_{\text{inc,2}}^{\text{TE/TM}}(k_y) \delta(k_x)
\end{align}
where $S^{\text{TE/TM}}(k_x,k_y)$ is the combination of two input signals in Fourier domain and $\delta(.)$ is Delta Dirac function. 
Based on the synthesized  OTF  (see \textcolor{blue}{Fig. 4}), the two-dimensional $O^{\text{TE/TM}}(k_x,k_y)$   can be written as follow
\begin{align}
& O^{\text{TE/TM}}(k_x,k_y)=\begin{cases}
\alpha j k_x & k_y=0 \\
\beta j  k_y & k_x=0 \\
W(k_x,k_y) & k_x,k_y \ne 0
\end{cases}
\end{align}
where $W(k_x,k_y)$ is corresponding to the synthesized OTF at ($k_x,k_y \ne 0$). \textcolor{black}{As we mentioned before, due to the fact that our input signals only have x or y variations, the synthesized OTF associated with metasurface processor  needs to have the first-order differentiator function only in $(k_x,0)$ and $(0,k_y)$. Therefore, the expression for $W(k_x,k_y)$ is unimportant for our purpose.} By multiplying the OTF to the $S^{\text{TE/TM}}(k_x,k_y)$, the output $V(k_x,k_y)$ can be written as follow  
\begin{align}
V(k_x,k_y) &= O^{\text{TE/TM}}(k_x,k_y) S^{\text{TE/TM}}(k_x,k_y) \\ &= \begin{cases}
\alpha j k_x \Psi_{\text{inc,1}}^{\text{TE/TM}}(k_x) \delta(k_y)  \\ +\alpha j k_x \Psi_{\text{inc,2}}^{\text{TE/TM}}(k_y) \delta(k_x) & k_y=0 \\\nonumber 
\beta j  k_y \Psi_{\text{inc,1}}^{\text{TE/TM}}(k_x) \delta(k_y) \\+ \beta j  k_y \Psi_{\text{inc,2}}^{\text{TE/TM}}(k_y) \delta(k_x) & k_x=0 \\ 
0 & k_x,k_y \ne 0
\end{cases} \\\nonumber
\end{align}

Based on the definition of delta Dirac function \cite{gelfand2} and after some mathematical manipulations we have 
$V(k_x,k_y)=\alpha j k_x \Psi_{\text{inc,1}}^{\text{TE/TM}}(k_x) \delta(k_y) + \beta j  k_y \Psi_{\text{inc,2}}^{\text{TE/TM}}(k_y) \delta(k_x)$; therefore, by using inverse Fourier Transform  $v(x,y)$  can be written as follow \cite{Youssefi:16,Zhu_2017}  

\begin{align}
&v(x,y) =\\
&\int_{-\infty}^{+\infty} \int_{-\infty}^{+\infty} \alpha j k_x \Psi_{\text{inc,1}}^{\text{TE/TM}}(k_x) \delta(k_y) \, e^{- j k_x x} e^{- j k_y y} \,dk_x\,dk_y \\ \nonumber
& + \int_{-\infty}^{+\infty} \int_{-\infty}^{+\infty} \beta j  k_y \Psi_{\text{inc,2}}^{\text{TE/TM}}(k_y) \delta(k_x) \, e^{- j k_x x} e^{- j k_y y} \,dk_x\,dk_y \\\nonumber
&= \alpha \frac{d}{dx}\psi_{\text{inc,1}}^{\text{TE/TM}}(x)+ \beta \frac{d}{dy}\psi_{\text{inc,2}}^{\text{TE/TM}}(y) 
\end{align}

\textcolor{black}{Finally, by using two SLPFs, we can extract and separate our desired output signals as following
\begin{align}
&k_y\text{-SPLF}\{v(x,y)\}=\alpha \frac{d}{dx}\psi_{\text{inc,1}}^{\text{TE/TM}}(x)\\
&k_x\text{-SPLF}\{v(x,y)\}= \beta \frac{d}{dy}\psi_{\text{inc,2}}^{\text{TE/TM}}(y) 
\end{align}}

Conceptually, the schematic sketch of the
2$\times$2 MIMO first-order differentiation processor is shown in \textcolor{blue}{Fig. 1b}.
Consider a computational and realistic scenario in which two different input signals in x and y directions ($\psi_{\text{inc,1}}^{\text{TE/TM}}(x)$ and $\psi_{\text{inc,2}}^{\text{TE/TM}}(y)$) collide to a reciprocal metasurface from normal direction. Upon interacting with the metasurface computer and depending on the $k_t$-modulation of the spatial OTF dictated by array of asymmetric meta-atoms, the combination of input signals ($s^{\text{TE/TM}}(x,y)$) is mapped  along  $k_x$ and $k_y$ directions of output signal ($V^{\text{TE/TM}}(k_x,k_y)$) in Fourier domain (see \textcolor{blue}{Equation.10}). Thereafter, desired output signals ($\frac{d}{dx}\psi_{\text{inc,1}}^{\text{TE/TM}}(x)$ and $\frac{d}{dy}\psi_{\text{inc,2}}^{\text{TE/TM}}(y) 
$) can be obtained and separated after passing from Spatial Low-Pass Filters (SLPF).  	

\begin{figure*}[t]%
	\centering
	{%
		\label{fig:3}%
		\includegraphics[scale=.6]{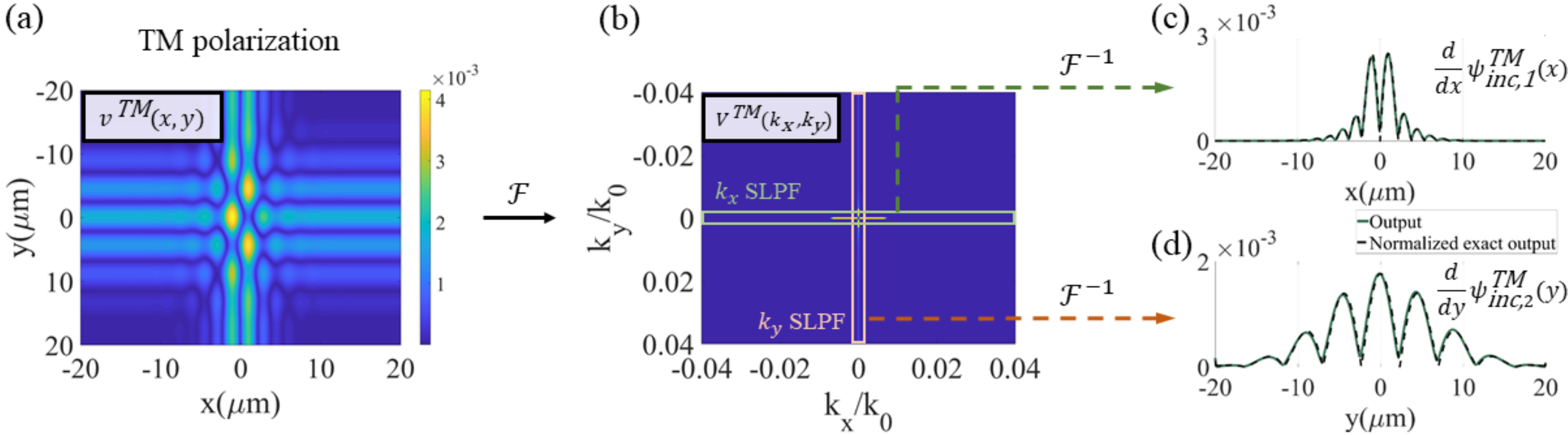}}%
	\caption{Output signals. TM-polarized transmitted image in space domain (a) and in Fourier domain (b). (c) and (d) are  extracted output signals after passing from $k_x$-SLPF and $k_y$-SLPF, respectively.  }
	\label{fig:5}
\end{figure*}
\begin{figure}[t!]%
	\centering
	{%
		\label{fig:3}%
		\includegraphics[scale=.27]{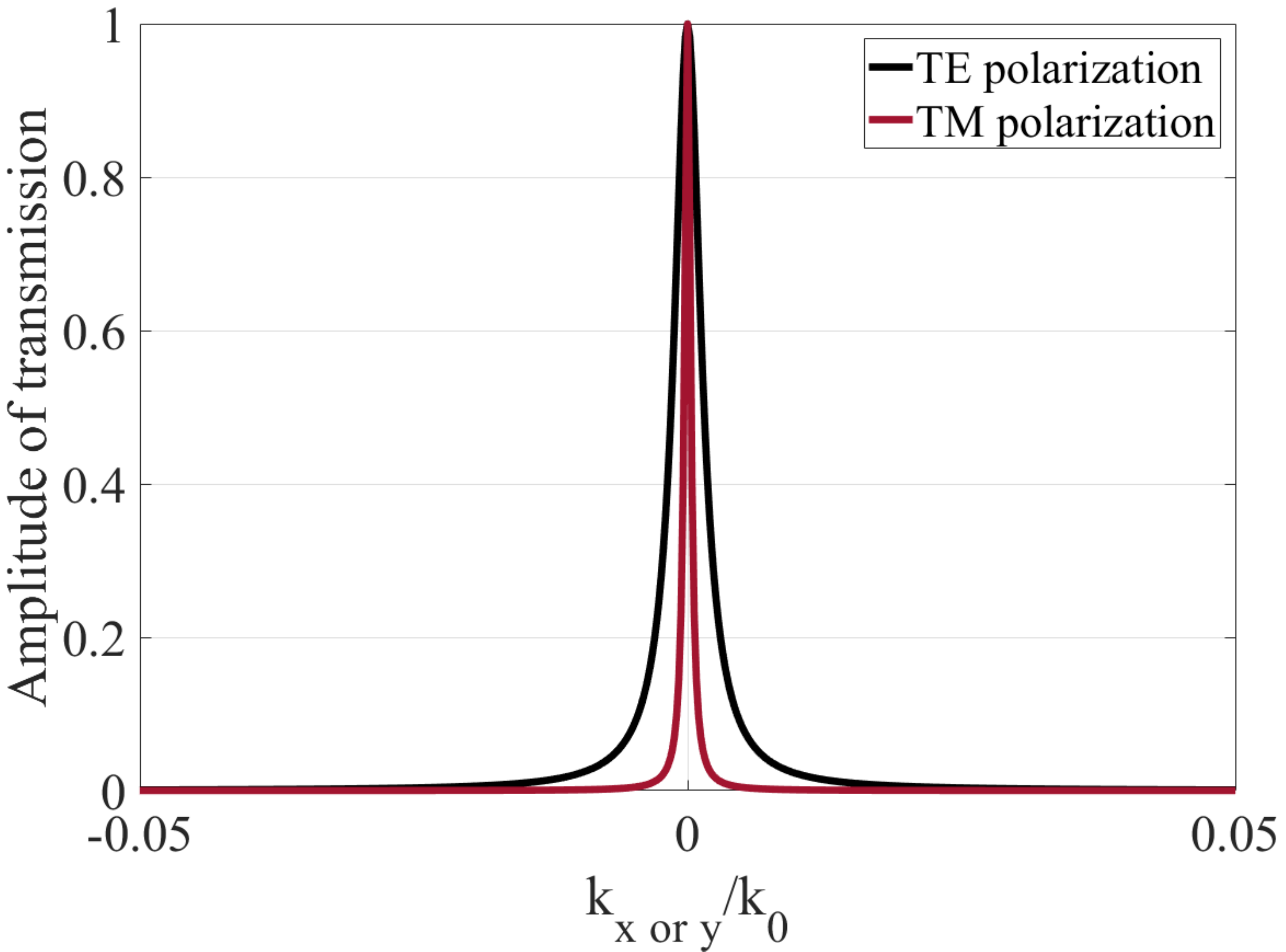}}%
	\caption{  Transmission coefficients of $k_x$- $k_y$-SLPF for both TE and TM states.}
	\label{fig:5}
\end{figure}
To demonstrate the real-time parallel computing capability of the proposed scheme, we consider as inputs the two arbitrary signals shown in \textcolor{blue}{Fig. 5a and b}. Simultaneously emitted by two sources, they combine into the two-dimensional function shown in \textcolor{blue}{Fig. 5c}. This field illuminates the metasurface computer, which generates the output image ($v^{\text{TE/TM}}(x,y)$) (see \textcolor{blue}{Figs. 6a and 7a}). Consistent with our design, the $V^{\text{TE/TM}}(k_x,k_y)$ image only has values along $k_x$ and $k_y$ directions (see \textcolor{blue}{Figs. 6b and 7b} and \textcolor{blue}{Equation. 10}. Subsequently, by using the SLPFs along $kx$ and $k_y$, the final output signals are generated and shown in \textcolor{blue}{Figs. 6 and 7}, for TE and TM polarizations, respectively. 
\textcolor{black}{One should note that many prior arts have introduced designs for appropriate SLPF based on a simple single metasurface \cite{PhysRevLett.121.173004,art22icle,zangeneh2017spatial}. The extraction performance of desired output signals depend on {these} SLPFs. Actually, the sharper {the} SLPFs, the better {the} extraction and separation {performance}. For this reason, we cascade two first-order integrator metasurface based on \cite{zangeneh2017spatial} to reach sharp-SLPFs. The transmission coefficients of desired SLPF are shown in \textcolor{blue}{Fig.8}.    }
For the sake of comparison, \textcolor{blue}{Figs. 6c and d } and \textcolor{blue}{Figs. 7c and d }also  plots the exact and the output signals for both orthogonal polarizations. 	An excellent agreement between simulated and benchmark results has been achieved when the normalized spectral beamwidth of the input signals lies within $|W|<0.35k_0$.
    

\section{conclusion}
~~~In summary, a 2$\times$2 MIMO first-order differentiation processor was elaborately
designed to be utilized in a spatial analog computing platform for reaching massively parallel processing.  Through synthesizing proper asymmetric meta-atom, based on GSTC and susceptibility tensors, the metasurface computer is	empowered to  realize the required phases and amplitudes of the OTF	associated with the first-order differentiation operation.  
This  novel optical MIMO metasurface processor with asymmetric optical response  can perform spatial differentiation on two distinct input signals (in x and y directions) at the same time for both orthogonal polarization states.	 The proposed theoretical framework foretastes that the presented design overcomes the substantial restrictions imposed by previous investigations such as large architectures arising from the need of additional subblocks, slow responses, and most importantly, supporting only one/single input and certain incident polarization.  The numerical results prove that the proposed metasurface computer may be thought of as an efficient and flexible host for being utilized in the field of real-time parallel optical signal processing.

\label{sec:refs}



\bibliography{sample}

\begin{thebibliography}{55}%
\makeatletter
\providecommand \@ifxundefined [1]{%
 \@ifx{#1\undefined}
}%
\providecommand \@ifnum [1]{%
 \ifnum #1\expandafter \@firstoftwo
 \else \expandafter \@secondoftwo
 \fi
}%
\providecommand \@ifx [1]{%
 \ifx #1\expandafter \@firstoftwo
 \else \expandafter \@secondoftwo
 \fi
}%
\providecommand \natexlab [1]{#1}%
\providecommand \enquote  [1]{``#1''}%
\providecommand \bibnamefont  [1]{#1}%
\providecommand \bibfnamefont [1]{#1}%
\providecommand \citenamefont [1]{#1}%
\providecommand \href@noop [0]{\@secondoftwo}%
\providecommand \href [0]{\begingroup \@sanitize@url \@href}%
\providecommand \@href[1]{\@@startlink{#1}\@@href}%
\providecommand \@@href[1]{\endgroup#1\@@endlink}%
\providecommand \@sanitize@url [0]{\catcode `\\12\catcode `\$12\catcode
  `\&12\catcode `\#12\catcode `\^12\catcode `\_12\catcode `\%12\relax}%
\providecommand \@@startlink[1]{}%
\providecommand \@@endlink[0]{}%
\providecommand \url  [0]{\begingroup\@sanitize@url \@url }%
\providecommand \@url [1]{\endgroup\@href {#1}{\urlprefix }}%
\providecommand \urlprefix  [0]{URL }%
\providecommand \Eprint [0]{\href }%
\providecommand \doibase [0]{http://dx.doi.org/}%
\providecommand \selectlanguage [0]{\@gobble}%
\providecommand \bibinfo  [0]{\@secondoftwo}%
\providecommand \bibfield  [0]{\@secondoftwo}%
\providecommand \translation [1]{[#1]}%
\providecommand \BibitemOpen [0]{}%
\providecommand \bibitemStop [0]{}%
\providecommand \bibitemNoStop [0]{.\EOS\space}%
\providecommand \EOS [0]{\spacefactor3000\relax}%
\providecommand \BibitemShut  [1]{\csname bibitem#1\endcsname}%
\let\auto@bib@innerbib\@empty
\bibitem [{\citenamefont {Nakamura}(2005)}]{white}%
  \BibitemOpen
  \bibfield  {author} {\bibinfo {author} {\bibfnamefont {J.}~\bibnamefont
  {Nakamura}},\ }\href@noop {} {\  (\bibinfo {year} {2005})}\BibitemShut
  {NoStop}%
\bibitem [{\citenamefont {Goodman}(2005)}]{goodman2005introduction}%
  \BibitemOpen
  \bibfield  {author} {\bibinfo {author} {\bibfnamefont {J.~W.}\ \bibnamefont
  {Goodman}},\ }\href@noop {} {\bibfield  {journal} {\bibinfo  {journal}
  {Introduction to Fourier optics, 3rd ed., by JW Goodman. Englewood, CO:
  Roberts \& Co. Publishers, 2005}\ }\textbf {\bibinfo {volume} {1}} (\bibinfo
  {year} {2005})}\BibitemShut {NoStop}%
\bibitem [{\citenamefont {Stark}(1982)}]{stark1982applications}%
  \BibitemOpen
  \bibfield  {author} {\bibinfo {author} {\bibfnamefont {H.}~\bibnamefont
  {Stark}},\ }\href {https://books.google.com/books?id=xvFRAAAAMAAJ} {\
  (\bibinfo {year} {1982})}\BibitemShut {NoStop}%
\bibitem [{\citenamefont {Goodman}(1990)}]{doi:10.1080/00043249.1990.10792698}%
  \BibitemOpen
  \bibfield  {author} {\bibinfo {author} {\bibfnamefont {C.}~\bibnamefont
  {Goodman}},\ }\href {\doibase 10.1080/00043249.1990.10792698} {\bibfield
  {journal} {\bibinfo  {journal} {Art Journal}\ }\textbf {\bibinfo {volume}
  {49}},\ \bibinfo {pages} {248} (\bibinfo {year} {1990})}\BibitemShut
  {NoStop}%
\bibitem [{\citenamefont {Silva}\ \emph {et~al.}(2014)\citenamefont {Silva},
  \citenamefont {Monticone}, \citenamefont {Castaldi}, \citenamefont {Galdi},
  \citenamefont {Al{\`u}},\ and\ \citenamefont {Engheta}}]{Silva160}%
  \BibitemOpen
  \bibfield  {author} {\bibinfo {author} {\bibfnamefont {A.}~\bibnamefont
  {Silva}}, \bibinfo {author} {\bibfnamefont {F.}~\bibnamefont {Monticone}},
  \bibinfo {author} {\bibfnamefont {G.}~\bibnamefont {Castaldi}}, \bibinfo
  {author} {\bibfnamefont {V.}~\bibnamefont {Galdi}}, \bibinfo {author}
  {\bibfnamefont {A.}~\bibnamefont {Al{\`u}}}, \ and\ \bibinfo {author}
  {\bibfnamefont {N.}~\bibnamefont {Engheta}},\ }\href {\doibase
  10.1126/science.1242818} {\bibfield  {journal} {\bibinfo  {journal}
  {Science}\ }\textbf {\bibinfo {volume} {343}},\ \bibinfo {pages} {160}
  (\bibinfo {year} {2014})}\BibitemShut {NoStop}%
\bibitem [{\citenamefont {{Solli}}\ and\ \citenamefont
  {{Jalali}}(2015)}]{2015NaPho...9..704S}%
  \BibitemOpen
  \bibfield  {author} {\bibinfo {author} {\bibfnamefont {D.~R.}\ \bibnamefont
  {{Solli}}}\ and\ \bibinfo {author} {\bibfnamefont {B.}~\bibnamefont
  {{Jalali}}},\ }\href {\doibase 10.1038/nphoton.2015.208} {\bibfield
  {journal} {\bibinfo  {journal} {Nature Photonics}\ }\textbf {\bibinfo
  {volume} {9}},\ \bibinfo {pages} {704} (\bibinfo {year} {2015})}\BibitemShut
  {NoStop}%
\bibitem [{\citenamefont {Abdollahramezani}\ \emph {et~al.}(2017)\citenamefont
  {Abdollahramezani}, \citenamefont {Chizari}, \citenamefont {Dorche},
  \citenamefont {Jamali},\ and\ \citenamefont
  {Salehi}}]{abdollahramezani2017dielectric}%
  \BibitemOpen
  \bibfield  {author} {\bibinfo {author} {\bibfnamefont {S.}~\bibnamefont
  {Abdollahramezani}}, \bibinfo {author} {\bibfnamefont {A.}~\bibnamefont
  {Chizari}}, \bibinfo {author} {\bibfnamefont {A.~E.}\ \bibnamefont {Dorche}},
  \bibinfo {author} {\bibfnamefont {M.~V.}\ \bibnamefont {Jamali}}, \ and\
  \bibinfo {author} {\bibfnamefont {J.~A.}\ \bibnamefont {Salehi}},\
  }\href@noop {} {\bibfield  {journal} {\bibinfo  {journal} {Optics letters}\
  }\textbf {\bibinfo {volume} {42}},\ \bibinfo {pages} {1197} (\bibinfo {year}
  {2017})}\BibitemShut {NoStop}%
\bibitem [{\citenamefont {Babashah}\ \emph {et~al.}(2017)\citenamefont
  {Babashah}, \citenamefont {Kavehvash}, \citenamefont {Koohi},\ and\
  \citenamefont {Khavasi}}]{Babashah:17}%
  \BibitemOpen
  \bibfield  {author} {\bibinfo {author} {\bibfnamefont {H.}~\bibnamefont
  {Babashah}}, \bibinfo {author} {\bibfnamefont {Z.}~\bibnamefont {Kavehvash}},
  \bibinfo {author} {\bibfnamefont {S.}~\bibnamefont {Koohi}}, \ and\ \bibinfo
  {author} {\bibfnamefont {A.}~\bibnamefont {Khavasi}},\ }\href@noop {}
  {\bibfield  {journal} {\bibinfo  {journal} {J. Opt. Soc. Am. B}\ }\textbf
  {\bibinfo {volume} {34}},\ \bibinfo {pages} {1270} (\bibinfo {year}
  {2017})}\BibitemShut {NoStop}%
\bibitem [{\citenamefont {Pors}\ \emph {et~al.}(2015)\citenamefont {Pors},
  \citenamefont {Nielsen},\ and\ \citenamefont {Bozhevolnyi}}]{pors2015analog}%
  \BibitemOpen
  \bibfield  {author} {\bibinfo {author} {\bibfnamefont {A.}~\bibnamefont
  {Pors}}, \bibinfo {author} {\bibfnamefont {M.~G.}\ \bibnamefont {Nielsen}}, \
  and\ \bibinfo {author} {\bibfnamefont {S.~I.}\ \bibnamefont {Bozhevolnyi}},\
  }\href@noop {} {\bibfield  {journal} {\bibinfo  {journal} {Nano letters}\
  }\textbf {\bibinfo {volume} {15}},\ \bibinfo {pages} {791} (\bibinfo {year}
  {2015})}\BibitemShut {NoStop}%
\bibitem [{\citenamefont {Youssefi}\ \emph {et~al.}(2016)\citenamefont
  {Youssefi}, \citenamefont {Zangeneh-Nejad}, \citenamefont
  {Abdollahramezani},\ and\ \citenamefont {Khavasi}}]{Youssefi:16}%
  \BibitemOpen
  \bibfield  {author} {\bibinfo {author} {\bibfnamefont {A.}~\bibnamefont
  {Youssefi}}, \bibinfo {author} {\bibfnamefont {F.}~\bibnamefont
  {Zangeneh-Nejad}}, \bibinfo {author} {\bibfnamefont {S.}~\bibnamefont
  {Abdollahramezani}}, \ and\ \bibinfo {author} {\bibfnamefont
  {A.}~\bibnamefont {Khavasi}},\ }\href@noop {} {\bibfield  {journal} {\bibinfo
   {journal} {Opt. Lett.}\ }\textbf {\bibinfo {volume} {41}},\ \bibinfo {pages}
  {3467} (\bibinfo {year} {2016})}\BibitemShut {NoStop}%
\bibitem [{\citenamefont {Kwon}\ \emph {et~al.}(2018)\citenamefont {Kwon},
  \citenamefont {Sounas}, \citenamefont {Cordaro}, \citenamefont {Polman},\
  and\ \citenamefont {Al\`u}}]{PhysRevLett.121.173004}%
  \BibitemOpen
  \bibfield  {author} {\bibinfo {author} {\bibfnamefont {H.}~\bibnamefont
  {Kwon}}, \bibinfo {author} {\bibfnamefont {D.}~\bibnamefont {Sounas}},
  \bibinfo {author} {\bibfnamefont {A.}~\bibnamefont {Cordaro}}, \bibinfo
  {author} {\bibfnamefont {A.}~\bibnamefont {Polman}}, \ and\ \bibinfo {author}
  {\bibfnamefont {A.}~\bibnamefont {Al\`u}},\ }\href@noop {} {\bibfield
  {journal} {\bibinfo  {journal} {Phys. Rev. Lett.}\ }\textbf {\bibinfo
  {volume} {121}},\ \bibinfo {pages} {173004} (\bibinfo {year}
  {2018})}\BibitemShut {NoStop}%
\bibitem [{ZAN(2018)}]{ZANGENEHNEJAD2018338}%
  \BibitemOpen
  \href@noop {} {\bibfield  {journal} {\bibinfo  {journal} {Optics
  Communications}\ }\textbf {\bibinfo {volume} {407}},\ \bibinfo {pages} {338 }
  (\bibinfo {year} {2018})}\BibitemShut {NoStop}%
\bibitem [{\citenamefont {Momeni}\ \emph
  {et~al.}(2019{\natexlab{a}})\citenamefont {Momeni}, \citenamefont
  {Rajabalipanah}, \citenamefont {Abdolali},\ and\ \citenamefont
  {Achouri}}]{PhysRevApplied.11.064042}%
  \BibitemOpen
  \bibfield  {author} {\bibinfo {author} {\bibfnamefont {A.}~\bibnamefont
  {Momeni}}, \bibinfo {author} {\bibfnamefont {H.}~\bibnamefont
  {Rajabalipanah}}, \bibinfo {author} {\bibfnamefont {A.}~\bibnamefont
  {Abdolali}}, \ and\ \bibinfo {author} {\bibfnamefont {K.}~\bibnamefont
  {Achouri}},\ }\href@noop {} {\bibfield  {journal} {\bibinfo  {journal} {Phys.
  Rev. Applied}\ }\textbf {\bibinfo {volume} {11}},\ \bibinfo {pages} {064042}
  (\bibinfo {year} {2019}{\natexlab{a}})}\BibitemShut {NoStop}%
\bibitem [{\citenamefont {Zangeneh-Nejad}\ and\ \citenamefont
  {Fleury}(2019)}]{Zangeneh_Nejad_2019}%
  \BibitemOpen
  \bibfield  {author} {\bibinfo {author} {\bibfnamefont {F.}~\bibnamefont
  {Zangeneh-Nejad}}\ and\ \bibinfo {author} {\bibfnamefont {R.}~\bibnamefont
  {Fleury}},\ }\href {\doibase 10.1038/s41467-019-10086-3} {\bibfield
  {journal} {\bibinfo  {journal} {Nature Communications}\ }\textbf {\bibinfo
  {volume} {10}} (\bibinfo {year} {2019}),\
  10.1038/s41467-019-10086-3}\BibitemShut {NoStop}%
\bibitem [{\citenamefont {Abdolali}\ \emph {et~al.}(2019)\citenamefont
  {Abdolali}, \citenamefont {Momeni}, \citenamefont {Rajabalipanah},\ and\
  \citenamefont {Achouri}}]{Abdolali:273146}%
  \BibitemOpen
  \bibfield  {author} {\bibinfo {author} {\bibfnamefont {A.}~\bibnamefont
  {Abdolali}}, \bibinfo {author} {\bibfnamefont {A.}~\bibnamefont {Momeni}},
  \bibinfo {author} {\bibfnamefont {H.}~\bibnamefont {Rajabalipanah}}, \ and\
  \bibinfo {author} {\bibfnamefont {K.}~\bibnamefont {Achouri}},\ }\href@noop
  {} {\bibfield  {journal} {\bibinfo  {journal} {New Journal Of Physics}\
  }\textbf {\bibinfo {volume} {21}},\ \bibinfo {pages} {113048} (\bibinfo
  {year} {2019})}\BibitemShut {NoStop}%
\bibitem [{\citenamefont {Zhu}\ \emph {et~al.}(2019)\citenamefont {Zhu},
  \citenamefont {Lou}, \citenamefont {Zhou}, \citenamefont {Zhang},
  \citenamefont {Huang}, \citenamefont {Li}, \citenamefont {Luo}, \citenamefont
  {Wen}, \citenamefont {Zhu}, \citenamefont {Gong}, \citenamefont {Qiu},\ and\
  \citenamefont {Ruan}}]{PhysRevApplied.11.034043}%
  \BibitemOpen
  \bibfield  {author} {\bibinfo {author} {\bibfnamefont {T.}~\bibnamefont
  {Zhu}}, \bibinfo {author} {\bibfnamefont {Y.}~\bibnamefont {Lou}}, \bibinfo
  {author} {\bibfnamefont {Y.}~\bibnamefont {Zhou}}, \bibinfo {author}
  {\bibfnamefont {J.}~\bibnamefont {Zhang}}, \bibinfo {author} {\bibfnamefont
  {J.}~\bibnamefont {Huang}}, \bibinfo {author} {\bibfnamefont
  {Y.}~\bibnamefont {Li}}, \bibinfo {author} {\bibfnamefont {H.}~\bibnamefont
  {Luo}}, \bibinfo {author} {\bibfnamefont {S.}~\bibnamefont {Wen}}, \bibinfo
  {author} {\bibfnamefont {S.}~\bibnamefont {Zhu}}, \bibinfo {author}
  {\bibfnamefont {Q.}~\bibnamefont {Gong}}, \bibinfo {author} {\bibfnamefont
  {M.}~\bibnamefont {Qiu}}, \ and\ \bibinfo {author} {\bibfnamefont
  {Z.}~\bibnamefont {Ruan}},\ }\href@noop {} {\bibfield  {journal} {\bibinfo
  {journal} {Phys. Rev. Applied}\ }\textbf {\bibinfo {volume} {11}},\ \bibinfo
  {pages} {034043} (\bibinfo {year} {2019})}\BibitemShut {NoStop}%
\bibitem [{\citenamefont {Zhu}\ \emph {et~al.}(2017)\citenamefont {Zhu},
  \citenamefont {Zhou}, \citenamefont {Lou}, \citenamefont {Ye}, \citenamefont
  {Qiu}, \citenamefont {Ruan},\ and\ \citenamefont {Fan}}]{Zhu_2017}%
  \BibitemOpen
  \bibfield  {author} {\bibinfo {author} {\bibfnamefont {T.}~\bibnamefont
  {Zhu}}, \bibinfo {author} {\bibfnamefont {Y.}~\bibnamefont {Zhou}}, \bibinfo
  {author} {\bibfnamefont {Y.}~\bibnamefont {Lou}}, \bibinfo {author}
  {\bibfnamefont {H.}~\bibnamefont {Ye}}, \bibinfo {author} {\bibfnamefont
  {M.}~\bibnamefont {Qiu}}, \bibinfo {author} {\bibfnamefont {Z.}~\bibnamefont
  {Ruan}}, \ and\ \bibinfo {author} {\bibfnamefont {S.}~\bibnamefont {Fan}},\
  }\href {\doibase 10.1038/ncomms15391} {\bibfield  {journal} {\bibinfo
  {journal} {Nature Communications}\ }\textbf {\bibinfo {volume} {8}} (\bibinfo
  {year} {2017}),\ 10.1038/ncomms15391}\BibitemShut {NoStop}%
\bibitem [{ZHO(2020)}]{ZHOU2020124674}%
  \BibitemOpen
  \href {\doibase https://doi.org/10.1016/j.optcom.2019.124674} {\bibfield
  {journal} {\bibinfo  {journal} {Optics Communications}\ }\textbf {\bibinfo
  {volume} {458}},\ \bibinfo {pages} {124674} (\bibinfo {year}
  {2020})}\BibitemShut {NoStop}%
\bibitem [{\citenamefont {Cordaro}\ \emph {et~al.}(2019)\citenamefont
  {Cordaro}, \citenamefont {Kwon}, \citenamefont {Sounas}, \citenamefont
  {Koenderink}, \citenamefont {Alù},\ and\ \citenamefont
  {Polman}}]{doi:10.1021/acs.nanolett.9b02477}%
  \BibitemOpen
  \bibfield  {author} {\bibinfo {author} {\bibfnamefont {A.}~\bibnamefont
  {Cordaro}}, \bibinfo {author} {\bibfnamefont {H.}~\bibnamefont {Kwon}},
  \bibinfo {author} {\bibfnamefont {D.}~\bibnamefont {Sounas}}, \bibinfo
  {author} {\bibfnamefont {A.~F.}\ \bibnamefont {Koenderink}}, \bibinfo
  {author} {\bibfnamefont {A.}~\bibnamefont {Alù}}, \ and\ \bibinfo {author}
  {\bibfnamefont {A.}~\bibnamefont {Polman}},\ }\href@noop {} {\bibfield
  {journal} {\bibinfo  {journal} {Nano Letters}\ }\textbf {\bibinfo {volume}
  {19}},\ \bibinfo {pages} {8418} (\bibinfo {year} {2019})}\BibitemShut
  {NoStop}%
\bibitem [{\citenamefont {Zhou}\ \emph {et~al.}(2020)\citenamefont {Zhou},
  \citenamefont {Wu}, \citenamefont {Chen}, \citenamefont {Chen}, \citenamefont
  {Chen},\ and\ \citenamefont {Ma}}]{doi:10.1002/adom.201901523}%
  \BibitemOpen
  \bibfield  {author} {\bibinfo {author} {\bibfnamefont {Y.}~\bibnamefont
  {Zhou}}, \bibinfo {author} {\bibfnamefont {W.}~\bibnamefont {Wu}}, \bibinfo
  {author} {\bibfnamefont {R.}~\bibnamefont {Chen}}, \bibinfo {author}
  {\bibfnamefont {W.}~\bibnamefont {Chen}}, \bibinfo {author} {\bibfnamefont
  {R.}~\bibnamefont {Chen}}, \ and\ \bibinfo {author} {\bibfnamefont
  {Y.}~\bibnamefont {Ma}},\ }\href@noop {} {\bibfield  {journal} {\bibinfo
  {journal} {Advanced Optical Materials}\ }\textbf {\bibinfo {volume} {8}},\
  \bibinfo {pages} {1901523} (\bibinfo {year} {2020})}\BibitemShut {NoStop}%
\bibitem [{\citenamefont {Mohammadi~Estakhri}\ \emph
  {et~al.}(2019)\citenamefont {Mohammadi~Estakhri}, \citenamefont {Edwards},\
  and\ \citenamefont {Engheta}}]{ss}%
  \BibitemOpen
  \bibfield  {author} {\bibinfo {author} {\bibfnamefont {N.}~\bibnamefont
  {Mohammadi~Estakhri}}, \bibinfo {author} {\bibfnamefont {B.}~\bibnamefont
  {Edwards}}, \ and\ \bibinfo {author} {\bibfnamefont {N.}~\bibnamefont
  {Engheta}},\ }\href@noop {} {\bibfield  {journal} {\bibinfo  {journal}
  {Science}\ }\textbf {\bibinfo {volume} {363}},\ \bibinfo {pages} {1333}
  (\bibinfo {year} {2019})}\BibitemShut {NoStop}%
\bibitem [{\citenamefont {Davis}\ \emph {et~al.}(2019)\citenamefont {Davis},
  \citenamefont {Eftekhari}, \citenamefont {G\'omez},\ and\ \citenamefont
  {Roberts}}]{PhysRevLett.123.013901}%
  \BibitemOpen
  \bibfield  {author} {\bibinfo {author} {\bibfnamefont {T.~J.}\ \bibnamefont
  {Davis}}, \bibinfo {author} {\bibfnamefont {F.}~\bibnamefont {Eftekhari}},
  \bibinfo {author} {\bibfnamefont {D.~E.}\ \bibnamefont {G\'omez}}, \ and\
  \bibinfo {author} {\bibfnamefont {A.}~\bibnamefont {Roberts}},\ }\href@noop
  {} {\bibfield  {journal} {\bibinfo  {journal} {Phys. Rev. Lett.}\ }\textbf
  {\bibinfo {volume} {123}},\ \bibinfo {pages} {013901} (\bibinfo {year}
  {2019})}\BibitemShut {NoStop}%
\bibitem [{\citenamefont {Guo}\ \emph {et~al.}(2018)\citenamefont {Guo},
  \citenamefont {Xiao}, \citenamefont {Minkov}, \citenamefont {Shi},\ and\
  \citenamefont {Fan}}]{Guo:18}%
  \BibitemOpen
  \bibfield  {author} {\bibinfo {author} {\bibfnamefont {C.}~\bibnamefont
  {Guo}}, \bibinfo {author} {\bibfnamefont {M.}~\bibnamefont {Xiao}}, \bibinfo
  {author} {\bibfnamefont {M.}~\bibnamefont {Minkov}}, \bibinfo {author}
  {\bibfnamefont {Y.}~\bibnamefont {Shi}}, \ and\ \bibinfo {author}
  {\bibfnamefont {S.}~\bibnamefont {Fan}},\ }\href@noop {} {\bibfield
  {journal} {\bibinfo  {journal} {Optica}\ }\textbf {\bibinfo {volume} {5}},\
  \bibinfo {pages} {251} (\bibinfo {year} {2018})}\BibitemShut {NoStop}%
\bibitem [{\citenamefont {Wang}\ \emph {et~al.}(2019)\citenamefont {Wang},
  \citenamefont {Li}, \citenamefont {Soman}, \citenamefont {Mao}, \citenamefont
  {Kananen},\ and\ \citenamefont {Gu}}]{PMID:31391468}%
  \BibitemOpen
  \bibfield  {author} {\bibinfo {author} {\bibfnamefont {Z.}~\bibnamefont
  {Wang}}, \bibinfo {author} {\bibfnamefont {T.}~\bibnamefont {Li}}, \bibinfo
  {author} {\bibfnamefont {A.}~\bibnamefont {Soman}}, \bibinfo {author}
  {\bibfnamefont {D.}~\bibnamefont {Mao}}, \bibinfo {author} {\bibfnamefont
  {T.}~\bibnamefont {Kananen}}, \ and\ \bibinfo {author} {\bibfnamefont
  {T.}~\bibnamefont {Gu}},\ }\href {\doibase 10.1038/s41467-019-11578-y}
  {\bibfield  {journal} {\bibinfo  {journal} {Nature communications}\ }\textbf
  {\bibinfo {volume} {10}},\ \bibinfo {pages} {3547} (\bibinfo {year}
  {2019})}\BibitemShut {NoStop}%
\bibitem [{\citenamefont {Karimi}\ \emph {et~al.}(2020)\citenamefont {Karimi},
  \citenamefont {Khavasi},\ and\ \citenamefont {Khaleghi}}]{Karimi:20}%
  \BibitemOpen
  \bibfield  {author} {\bibinfo {author} {\bibfnamefont {P.}~\bibnamefont
  {Karimi}}, \bibinfo {author} {\bibfnamefont {A.}~\bibnamefont {Khavasi}}, \
  and\ \bibinfo {author} {\bibfnamefont {S.~S.~M.}\ \bibnamefont {Khaleghi}},\
  }\href@noop {} {\bibfield  {journal} {\bibinfo  {journal} {Opt. Express}\
  }\textbf {\bibinfo {volume} {28}},\ \bibinfo {pages} {898} (\bibinfo {year}
  {2020})}\BibitemShut {NoStop}%
\bibitem [{\citenamefont {Zhou}\ \emph {et~al.}(2019)\citenamefont {Zhou},
  \citenamefont {Qian}, \citenamefont {Chen}, \citenamefont {Zhao},
  \citenamefont {Li}, \citenamefont {Wu}, \citenamefont {Luo}, \citenamefont
  {Wen},\ and\ \citenamefont {Liu}}]{Zhou11137}%
  \BibitemOpen
  \bibfield  {author} {\bibinfo {author} {\bibfnamefont {J.}~\bibnamefont
  {Zhou}}, \bibinfo {author} {\bibfnamefont {H.}~\bibnamefont {Qian}}, \bibinfo
  {author} {\bibfnamefont {C.-F.}\ \bibnamefont {Chen}}, \bibinfo {author}
  {\bibfnamefont {J.}~\bibnamefont {Zhao}}, \bibinfo {author} {\bibfnamefont
  {G.}~\bibnamefont {Li}}, \bibinfo {author} {\bibfnamefont {Q.}~\bibnamefont
  {Wu}}, \bibinfo {author} {\bibfnamefont {H.}~\bibnamefont {Luo}}, \bibinfo
  {author} {\bibfnamefont {S.}~\bibnamefont {Wen}}, \ and\ \bibinfo {author}
  {\bibfnamefont {Z.}~\bibnamefont {Liu}},\ }\href {\doibase
  10.1073/pnas.1820636116} {\bibfield  {journal} {\bibinfo  {journal}
  {Proceedings of the National Academy of Sciences}\ }\textbf {\bibinfo
  {volume} {116}},\ \bibinfo {pages} {11137} (\bibinfo {year}
  {2019})}\BibitemShut {NoStop}%
\bibitem [{\citenamefont {Rajabalipanah}\ \emph {et~al.}(2020)\citenamefont
  {Rajabalipanah}, \citenamefont {Abdolali}, \citenamefont {Iqbal},
  \citenamefont {Zhang},\ and\ \citenamefont
  {Cui}}]{rajabalipanah2020spacetime}%
  \BibitemOpen
  \bibfield  {author} {\bibinfo {author} {\bibfnamefont {H.}~\bibnamefont
  {Rajabalipanah}}, \bibinfo {author} {\bibfnamefont {A.}~\bibnamefont
  {Abdolali}}, \bibinfo {author} {\bibfnamefont {S.}~\bibnamefont {Iqbal}},
  \bibinfo {author} {\bibfnamefont {L.}~\bibnamefont {Zhang}}, \ and\ \bibinfo
  {author} {\bibfnamefont {T.~J.}\ \bibnamefont {Cui}},\ }\href@noop {} {\
  (\bibinfo {year} {2020})},\ \Eprint {http://arxiv.org/abs/2002.06773}
  {arXiv:2002.06773 [physics.app-ph]} \BibitemShut {NoStop}%
\bibitem [{\citenamefont {Yu}\ and\ \citenamefont
  {Capasso}(2014)}]{Yu2014FlatOW}%
  \BibitemOpen
  \bibfield  {author} {\bibinfo {author} {\bibfnamefont {N.}~\bibnamefont
  {Yu}}\ and\ \bibinfo {author} {\bibfnamefont {F.}~\bibnamefont {Capasso}},\
  }\href@noop {} {\bibfield  {journal} {\bibinfo  {journal} {Nature materials}\
  }\textbf {\bibinfo {volume} {13 2}},\ \bibinfo {pages} {139} (\bibinfo {year}
  {2014})}\BibitemShut {NoStop}%
\bibitem [{\citenamefont {Yu}\ \emph {et~al.}(2011)\citenamefont {Yu},
  \citenamefont {Genevet}, \citenamefont {Kats}, \citenamefont {Aieta},
  \citenamefont {Tetienne}, \citenamefont {Capasso},\ and\ \citenamefont
  {Gaburro}}]{Yu333}%
  \BibitemOpen
  \bibfield  {author} {\bibinfo {author} {\bibfnamefont {N.}~\bibnamefont
  {Yu}}, \bibinfo {author} {\bibfnamefont {P.}~\bibnamefont {Genevet}},
  \bibinfo {author} {\bibfnamefont {M.~A.}\ \bibnamefont {Kats}}, \bibinfo
  {author} {\bibfnamefont {F.}~\bibnamefont {Aieta}}, \bibinfo {author}
  {\bibfnamefont {J.-P.}\ \bibnamefont {Tetienne}}, \bibinfo {author}
  {\bibfnamefont {F.}~\bibnamefont {Capasso}}, \ and\ \bibinfo {author}
  {\bibfnamefont {Z.}~\bibnamefont {Gaburro}},\ }\href@noop {} {\bibfield
  {journal} {\bibinfo  {journal} {Science}\ }\textbf {\bibinfo {volume}
  {334}},\ \bibinfo {pages} {333} (\bibinfo {year} {2011})}\BibitemShut
  {NoStop}%
\bibitem [{\citenamefont {Sun}\ \emph {et~al.}(2012)\citenamefont {Sun},
  \citenamefont {Yang}, \citenamefont {Wang}, \citenamefont {Juan},
  \citenamefont {Chen}, \citenamefont {Liao}, \citenamefont {He}, \citenamefont
  {Xiao}, \citenamefont {Kung}, \citenamefont {Guo}, \citenamefont {Zhou},\
  and\ \citenamefont {Tsai}}]{doi:10.1021/nl3032668}%
  \BibitemOpen
  \bibfield  {author} {\bibinfo {author} {\bibfnamefont {S.}~\bibnamefont
  {Sun}}, \bibinfo {author} {\bibfnamefont {K.-Y.}\ \bibnamefont {Yang}},
  \bibinfo {author} {\bibfnamefont {C.-M.}\ \bibnamefont {Wang}}, \bibinfo
  {author} {\bibfnamefont {T.-K.}\ \bibnamefont {Juan}}, \bibinfo {author}
  {\bibfnamefont {W.~T.}\ \bibnamefont {Chen}}, \bibinfo {author}
  {\bibfnamefont {C.~Y.}\ \bibnamefont {Liao}}, \bibinfo {author}
  {\bibfnamefont {Q.}~\bibnamefont {He}}, \bibinfo {author} {\bibfnamefont
  {S.}~\bibnamefont {Xiao}}, \bibinfo {author} {\bibfnamefont {W.-T.}\
  \bibnamefont {Kung}}, \bibinfo {author} {\bibfnamefont {G.-Y.}\ \bibnamefont
  {Guo}}, \bibinfo {author} {\bibfnamefont {L.}~\bibnamefont {Zhou}}, \ and\
  \bibinfo {author} {\bibfnamefont {D.~P.}\ \bibnamefont {Tsai}},\ }\href
  {\doibase 10.1021/nl3032668} {\bibfield  {journal} {\bibinfo  {journal} {Nano
  Letters}\ }\textbf {\bibinfo {volume} {12}},\ \bibinfo {pages} {6223}
  (\bibinfo {year} {2012})},\ \bibinfo {note} {pMID: 23189928}\BibitemShut
  {NoStop}%
\bibitem [{\citenamefont {Kildishev}\ \emph {et~al.}(2013)\citenamefont
  {Kildishev}, \citenamefont {Boltasseva},\ and\ \citenamefont
  {Shalaev}}]{Kildishev1232009}%
  \BibitemOpen
  \bibfield  {author} {\bibinfo {author} {\bibfnamefont {A.~V.}\ \bibnamefont
  {Kildishev}}, \bibinfo {author} {\bibfnamefont {A.}~\bibnamefont
  {Boltasseva}}, \ and\ \bibinfo {author} {\bibfnamefont {V.~M.}\ \bibnamefont
  {Shalaev}},\ }\href {\doibase 10.1126/science.1232009} {\bibfield  {journal}
  {\bibinfo  {journal} {Science}\ }\textbf {\bibinfo {volume} {339}} (\bibinfo
  {year} {2013}),\ 10.1126/science.1232009}\BibitemShut {NoStop}%
\bibitem [{\citenamefont {{Achouri}}\ and\ \citenamefont
  {{Martin}}(2020)}]{8852838}%
  \BibitemOpen
  \bibfield  {author} {\bibinfo {author} {\bibfnamefont {K.}~\bibnamefont
  {{Achouri}}}\ and\ \bibinfo {author} {\bibfnamefont {O.~J.~F.}\ \bibnamefont
  {{Martin}}},\ }\href {\doibase 10.1109/TAP.2019.2943423} {\bibfield
  {journal} {\bibinfo  {journal} {IEEE Transactions on Antennas and
  Propagation}\ }\textbf {\bibinfo {volume} {68}},\ \bibinfo {pages} {432}
  (\bibinfo {year} {2020})}\BibitemShut {NoStop}%
\bibitem [{\citenamefont {Momeni}\ \emph
  {et~al.}(2019{\natexlab{b}})\citenamefont {Momeni}, \citenamefont {Safari},
  \citenamefont {Abdolali},\ and\ \citenamefont {Kherani}}]{momeni2019tunable}%
  \BibitemOpen
  \bibfield  {author} {\bibinfo {author} {\bibfnamefont {A.}~\bibnamefont
  {Momeni}}, \bibinfo {author} {\bibfnamefont {M.}~\bibnamefont {Safari}},
  \bibinfo {author} {\bibfnamefont {A.}~\bibnamefont {Abdolali}}, \ and\
  \bibinfo {author} {\bibfnamefont {N.~P.}\ \bibnamefont {Kherani}},\
  }\href@noop {} {\enquote {\bibinfo {title} {Tunable and dynamic
  polarizability tensor for asymmetric metal-dielectric meta-cylinders},}\ }
  (\bibinfo {year} {2019}{\natexlab{b}}),\ \Eprint
  {http://arxiv.org/abs/1904.04102} {arXiv:1904.04102 [physics.app-ph]}
  \BibitemShut {NoStop}%
\bibitem [{\citenamefont {Rouhi}\ \emph {et~al.}(2019)\citenamefont {Rouhi},
  \citenamefont {Rajabalipanah},\ and\ \citenamefont
  {Abdolali}}]{ROUHI2019125}%
  \BibitemOpen
  \bibfield  {author} {\bibinfo {author} {\bibfnamefont {K.}~\bibnamefont
  {Rouhi}}, \bibinfo {author} {\bibfnamefont {H.}~\bibnamefont
  {Rajabalipanah}}, \ and\ \bibinfo {author} {\bibfnamefont {A.}~\bibnamefont
  {Abdolali}},\ }\href@noop {} {\bibfield  {journal} {\bibinfo  {journal}
  {Carbon}\ }\textbf {\bibinfo {volume} {149}},\ \bibinfo {pages} {125 }
  (\bibinfo {year} {2019})}\BibitemShut {NoStop}%
\bibitem [{\citenamefont {Shahid}\ \emph {et~al.}(2020)\citenamefont {Shahid},
  \citenamefont {Rajabalipanah}, \citenamefont {Zhang}, \citenamefont {Qiang},
  \citenamefont {Abdolali},\ and\ \citenamefont {Cui}}]{Shahid2020}%
  \BibitemOpen
  \bibfield  {author} {\bibinfo {author} {\bibfnamefont {I.}~\bibnamefont
  {Shahid}}, \bibinfo {author} {\bibfnamefont {H.}~\bibnamefont
  {Rajabalipanah}}, \bibinfo {author} {\bibfnamefont {L.}~\bibnamefont
  {Zhang}}, \bibinfo {author} {\bibfnamefont {X.}~\bibnamefont {Qiang}},
  \bibinfo {author} {\bibfnamefont {A.}~\bibnamefont {Abdolali}}, \ and\
  \bibinfo {author} {\bibfnamefont {T.~J.}\ \bibnamefont {Cui}},\ }\href@noop
  {} {\bibfield  {journal} {\bibinfo  {journal} {Nanophotonics}\ }\textbf
  {\bibinfo {volume} {9}},\ \bibinfo {pages} {703} (\bibinfo {year} {2020})},\
  \bibinfo {note} {3}\BibitemShut {NoStop}%
\bibitem [{\citenamefont {Kiani}\ \emph {et~al.}(2020)\citenamefont {Kiani},
  \citenamefont {Tayarani}, \citenamefont {Momeni}, \citenamefont
  {Rajabalipanah},\ and\ \citenamefont {Abdolali}}]{Kiani:20}%
  \BibitemOpen
  \bibfield  {author} {\bibinfo {author} {\bibfnamefont {M.}~\bibnamefont
  {Kiani}}, \bibinfo {author} {\bibfnamefont {M.}~\bibnamefont {Tayarani}},
  \bibinfo {author} {\bibfnamefont {A.}~\bibnamefont {Momeni}}, \bibinfo
  {author} {\bibfnamefont {H.}~\bibnamefont {Rajabalipanah}}, \ and\ \bibinfo
  {author} {\bibfnamefont {A.}~\bibnamefont {Abdolali}},\ }\href@noop {}
  {\bibfield  {journal} {\bibinfo  {journal} {Opt. Express}\ }\textbf {\bibinfo
  {volume} {28}},\ \bibinfo {pages} {5410} (\bibinfo {year}
  {2020})}\BibitemShut {NoStop}%
\bibitem [{\citenamefont {Rouhi}\ \emph {et~al.}(2018)\citenamefont {Rouhi},
  \citenamefont {Rajabalipanah},\ and\ \citenamefont
  {Abdolali}}]{doi:10.1002/andp.201700310}%
  \BibitemOpen
  \bibfield  {author} {\bibinfo {author} {\bibfnamefont {K.}~\bibnamefont
  {Rouhi}}, \bibinfo {author} {\bibfnamefont {H.}~\bibnamefont
  {Rajabalipanah}}, \ and\ \bibinfo {author} {\bibfnamefont {A.}~\bibnamefont
  {Abdolali}},\ }\href@noop {} {\bibfield  {journal} {\bibinfo  {journal}
  {Annalen der Physik}\ }\textbf {\bibinfo {volume} {530}},\ \bibinfo {pages}
  {1700310} (\bibinfo {year} {2018})}\BibitemShut {NoStop}%
\bibitem [{\citenamefont {Rajabalipanah}\ \emph {et~al.}(2019)\citenamefont
  {Rajabalipanah}, \citenamefont {Abdolali}, \citenamefont {Shabanpour},
  \citenamefont {Momeni},\ and\ \citenamefont
  {Cheldavi}}]{doi:10.1021/acsomega.9b02195}%
  \BibitemOpen
  \bibfield  {author} {\bibinfo {author} {\bibfnamefont {H.}~\bibnamefont
  {Rajabalipanah}}, \bibinfo {author} {\bibfnamefont {A.}~\bibnamefont
  {Abdolali}}, \bibinfo {author} {\bibfnamefont {J.}~\bibnamefont
  {Shabanpour}}, \bibinfo {author} {\bibfnamefont {A.}~\bibnamefont {Momeni}},
  \ and\ \bibinfo {author} {\bibfnamefont {A.}~\bibnamefont {Cheldavi}},\
  }\href@noop {} {\bibfield  {journal} {\bibinfo  {journal} {ACS Omega}\
  }\textbf {\bibinfo {volume} {4}},\ \bibinfo {pages} {14340} (\bibinfo {year}
  {2019})}\BibitemShut {NoStop}%
\bibitem [{\citenamefont {Hosseininejad}\ \emph
  {et~al.}(2019{\natexlab{a}})\citenamefont {Hosseininejad}, \citenamefont
  {Rouhi}, \citenamefont {Neshat}, \citenamefont {Faraji-Dana}, \citenamefont
  {Cabellos-Aparicio}, \citenamefont {Abadal},\ and\ \citenamefont
  {Alarc{\'o}n}}]{Hosseininejad2019ReprogrammableGM}%
  \BibitemOpen
  \bibfield  {author} {\bibinfo {author} {\bibfnamefont {S.~E.}\ \bibnamefont
  {Hosseininejad}}, \bibinfo {author} {\bibfnamefont {K.}~\bibnamefont
  {Rouhi}}, \bibinfo {author} {\bibfnamefont {M.}~\bibnamefont {Neshat}},
  \bibinfo {author} {\bibfnamefont {R.}~\bibnamefont {Faraji-Dana}}, \bibinfo
  {author} {\bibfnamefont {A.}~\bibnamefont {Cabellos-Aparicio}}, \bibinfo
  {author} {\bibfnamefont {S.}~\bibnamefont {Abadal}}, \ and\ \bibinfo {author}
  {\bibfnamefont {E.}~\bibnamefont {Alarc{\'o}n}},\ }\bibfield  {booktitle}
  {\emph {\bibinfo {booktitle} {Scientific Reports}},\ }\href@noop {} {\
  (\bibinfo {year} {2019}{\natexlab{a}})}\BibitemShut {NoStop}%
\bibitem [{\citenamefont {Kargar}\ \emph {et~al.}(2020)\citenamefont {Kargar},
  \citenamefont {Rouhi},\ and\ \citenamefont {Abdolali}}]{KARGAR2020125331}%
  \BibitemOpen
  \bibfield  {author} {\bibinfo {author} {\bibfnamefont {R.}~\bibnamefont
  {Kargar}}, \bibinfo {author} {\bibfnamefont {K.}~\bibnamefont {Rouhi}}, \
  and\ \bibinfo {author} {\bibfnamefont {A.}~\bibnamefont {Abdolali}},\
  }\href@noop {} {\bibfield  {journal} {\bibinfo  {journal} {Optics
  Communications}\ }\textbf {\bibinfo {volume} {462}},\ \bibinfo {pages}
  {125331} (\bibinfo {year} {2020})}\BibitemShut {NoStop}%
\bibitem [{\citenamefont {Momeni}\ \emph {et~al.}(2018)\citenamefont {Momeni},
  \citenamefont {Rouhi}, \citenamefont {Rajabalipanah},\ and\ \citenamefont
  {Abdolali}}]{article1}%
  \BibitemOpen
  \bibfield  {author} {\bibinfo {author} {\bibfnamefont {A.}~\bibnamefont
  {Momeni}}, \bibinfo {author} {\bibfnamefont {K.}~\bibnamefont {Rouhi}},
  \bibinfo {author} {\bibfnamefont {H.}~\bibnamefont {Rajabalipanah}}, \ and\
  \bibinfo {author} {\bibfnamefont {A.}~\bibnamefont {Abdolali}},\ }\href
  {\doibase 10.1038/s41598-018-24553-2} {\bibfield  {journal} {\bibinfo
  {journal} {Scientific Reports}\ }\textbf {\bibinfo {volume} {8}} (\bibinfo
  {year} {2018}),\ 10.1038/s41598-018-24553-2}\BibitemShut {NoStop}%
\bibitem [{\citenamefont {Hosseininejad}\ \emph
  {et~al.}(2019{\natexlab{b}})\citenamefont {Hosseininejad}, \citenamefont
  {Rouhi}, \citenamefont {Neshat}, \citenamefont {Cabellos-Aparicio},
  \citenamefont {Abadal},\ and\ \citenamefont
  {Alarc{\'o}n}}]{Hosseininejad2019DigitalMB}%
  \BibitemOpen
  \bibfield  {author} {\bibinfo {author} {\bibfnamefont {S.~E.}\ \bibnamefont
  {Hosseininejad}}, \bibinfo {author} {\bibfnamefont {K.}~\bibnamefont
  {Rouhi}}, \bibinfo {author} {\bibfnamefont {M.}~\bibnamefont {Neshat}},
  \bibinfo {author} {\bibfnamefont {A.}~\bibnamefont {Cabellos-Aparicio}},
  \bibinfo {author} {\bibfnamefont {S.}~\bibnamefont {Abadal}}, \ and\ \bibinfo
  {author} {\bibfnamefont {E.}~\bibnamefont {Alarc{\'o}n}},\ }\href@noop {}
  {\bibfield  {journal} {\bibinfo  {journal} {IEEE Transactions on
  Nanotechnology}\ }\textbf {\bibinfo {volume} {18}},\ \bibinfo {pages} {734}
  (\bibinfo {year} {2019}{\natexlab{b}})}\BibitemShut {NoStop}%
\bibitem [{\citenamefont {Moeini}\ \emph
  {et~al.}(2019{\natexlab{a}})\citenamefont {Moeini}, \citenamefont {Oraizi},\
  and\ \citenamefont {Amini}}]{PhysRevApplied.11.044006}%
  \BibitemOpen
  \bibfield  {author} {\bibinfo {author} {\bibfnamefont {M.~M.}\ \bibnamefont
  {Moeini}}, \bibinfo {author} {\bibfnamefont {H.}~\bibnamefont {Oraizi}}, \
  and\ \bibinfo {author} {\bibfnamefont {A.}~\bibnamefont {Amini}},\
  }\href@noop {} {\bibfield  {journal} {\bibinfo  {journal} {Phys. Rev.
  Applied}\ }\textbf {\bibinfo {volume} {11}},\ \bibinfo {pages} {044006}
  (\bibinfo {year} {2019}{\natexlab{a}})}\BibitemShut {NoStop}%
\bibitem [{\citenamefont {{Movahhedi}}\ \emph {et~al.}(2019)\citenamefont
  {{Movahhedi}}, \citenamefont {{Karimipour}},\ and\ \citenamefont
  {{Komjani}}}]{8731665}%
  \BibitemOpen
  \bibfield  {author} {\bibinfo {author} {\bibfnamefont {M.}~\bibnamefont
  {{Movahhedi}}}, \bibinfo {author} {\bibfnamefont {M.}~\bibnamefont
  {{Karimipour}}}, \ and\ \bibinfo {author} {\bibfnamefont {N.}~\bibnamefont
  {{Komjani}}},\ }\href@noop {} {\bibfield  {journal} {\bibinfo  {journal}
  {IEEE Antennas and Wireless Propagation Letters}\ }\textbf {\bibinfo {volume}
  {18}},\ \bibinfo {pages} {1507} (\bibinfo {year} {2019})}\BibitemShut
  {NoStop}%
\bibitem [{\citenamefont {Moeini}\ \emph
  {et~al.}(2019{\natexlab{b}})\citenamefont {Moeini}, \citenamefont {Oraizi},
  \citenamefont {Amini},\ and\ \citenamefont {Nayyeri}}]{article}%
  \BibitemOpen
  \bibfield  {author} {\bibinfo {author} {\bibfnamefont {M.~M.}\ \bibnamefont
  {Moeini}}, \bibinfo {author} {\bibfnamefont {H.}~\bibnamefont {Oraizi}},
  \bibinfo {author} {\bibfnamefont {A.}~\bibnamefont {Amini}}, \ and\ \bibinfo
  {author} {\bibfnamefont {V.}~\bibnamefont {Nayyeri}},\ }\href@noop {}
  {\bibfield  {journal} {\bibinfo  {journal} {Scientific Reports}\ }\textbf
  {\bibinfo {volume} {9}} (\bibinfo {year} {2019}{\natexlab{b}})}\BibitemShut
  {NoStop}%
\bibitem [{\citenamefont {{Canny}}(1986)}]{4767851}%
  \BibitemOpen
  \bibfield  {author} {\bibinfo {author} {\bibfnamefont {J.}~\bibnamefont
  {{Canny}}},\ }\href@noop {} {\bibfield  {journal} {\bibinfo  {journal} {IEEE
  Transactions on Pattern Analysis and Machine Intelligence}\ }\textbf
  {\bibinfo {volume} {PAMI-8}},\ \bibinfo {pages} {679} (\bibinfo {year}
  {1986})}\BibitemShut {NoStop}%
\bibitem [{\citenamefont {Marr}\ \emph {et~al.}(1980)\citenamefont {Marr},
  \citenamefont {Hildreth},\ and\ \citenamefont
  {Brenner}}]{doi:10.1098/rspb.1980.0020}%
  \BibitemOpen
  \bibfield  {author} {\bibinfo {author} {\bibfnamefont {D.}~\bibnamefont
  {Marr}}, \bibinfo {author} {\bibfnamefont {E.}~\bibnamefont {Hildreth}}, \
  and\ \bibinfo {author} {\bibfnamefont {S.}~\bibnamefont {Brenner}},\
  }\href@noop {} {\bibfield  {journal} {\bibinfo  {journal} {Proceedings of the
  Royal Society of London. Series B. Biological Sciences}\ }\textbf {\bibinfo
  {volume} {207}},\ \bibinfo {pages} {187} (\bibinfo {year}
  {1980})}\BibitemShut {NoStop}%
\bibitem [{\citenamefont {Achouri}\ \emph {et~al.}(2016)\citenamefont
  {Achouri}, \citenamefont {Lavigne},\ and\ \citenamefont
  {Caloz}}]{doi:10.1063/1.4972195}%
  \BibitemOpen
  \bibfield  {author} {\bibinfo {author} {\bibfnamefont {K.}~\bibnamefont
  {Achouri}}, \bibinfo {author} {\bibfnamefont {G.}~\bibnamefont {Lavigne}}, \
  and\ \bibinfo {author} {\bibfnamefont {C.}~\bibnamefont {Caloz}},\
  }\href@noop {} {\bibfield  {journal} {\bibinfo  {journal} {Journal of Applied
  Physics}\ }\textbf {\bibinfo {volume} {120}},\ \bibinfo {pages} {235305}
  (\bibinfo {year} {2016})}\BibitemShut {NoStop}%
\bibitem [{\citenamefont {Achouri}\ and\ \citenamefont
  {Caloz}(2017)}]{art3icle}%
  \BibitemOpen
  \bibfield  {author} {\bibinfo {author} {\bibfnamefont {K.}~\bibnamefont
  {Achouri}}\ and\ \bibinfo {author} {\bibfnamefont {C.}~\bibnamefont
  {Caloz}},\ }\href@noop {} {\bibfield  {journal} {\bibinfo  {journal}
  {Nanophotonics}\ } (\bibinfo {year} {2017})}\BibitemShut {NoStop}%
\bibitem [{\citenamefont {{Achouri, Karim}}\ \emph {et~al.}(2015)\citenamefont
  {{Achouri, Karim}}, \citenamefont {{Khan, Bakthiar Ali}}, \citenamefont
  {{Gupta, Shulabh}}, \citenamefont {{Lavigne, Guillaume}}, \citenamefont
  {{Salem, Mohamed Ahmed}},\ and\ \citenamefont {{Caloz,
  Christophe}}}]{refId0}%
  \BibitemOpen
  \bibfield  {author} {\bibinfo {author} {\bibnamefont {{Achouri, Karim}}},
  \bibinfo {author} {\bibnamefont {{Khan, Bakthiar Ali}}}, \bibinfo {author}
  {\bibnamefont {{Gupta, Shulabh}}}, \bibinfo {author} {\bibnamefont {{Lavigne,
  Guillaume}}}, \bibinfo {author} {\bibnamefont {{Salem, Mohamed Ahmed}}}, \
  and\ \bibinfo {author} {\bibnamefont {{Caloz, Christophe}}},\ }\href@noop {}
  {\bibfield  {journal} {\bibinfo  {journal} {EPJ Applied Metamaterials}\
  }\textbf {\bibinfo {volume} {2}},\ \bibinfo {pages} {12} (\bibinfo {year}
  {2015})}\BibitemShut {NoStop}%
\bibitem [{\citenamefont {{Lavigne}}\ \emph {et~al.}(2018)\citenamefont
  {{Lavigne}}, \citenamefont {{Achouri}}, \citenamefont {{Asadchy}},
  \citenamefont {{Tretyakov}},\ and\ \citenamefont {{Caloz}}}]{8259235}%
  \BibitemOpen
  \bibfield  {author} {\bibinfo {author} {\bibfnamefont {G.}~\bibnamefont
  {{Lavigne}}}, \bibinfo {author} {\bibfnamefont {K.}~\bibnamefont
  {{Achouri}}}, \bibinfo {author} {\bibfnamefont {V.~S.}\ \bibnamefont
  {{Asadchy}}}, \bibinfo {author} {\bibfnamefont {S.~A.}\ \bibnamefont
  {{Tretyakov}}}, \ and\ \bibinfo {author} {\bibfnamefont {C.}~\bibnamefont
  {{Caloz}}},\ }\href@noop {} {\bibfield  {journal} {\bibinfo  {journal} {IEEE
  Transactions on Antennas and Propagation}\ }\textbf {\bibinfo {volume}
  {66}},\ \bibinfo {pages} {1321} (\bibinfo {year} {2018})}\BibitemShut
  {NoStop}%
\bibitem [{\citenamefont {Palik}(1998)}]{hand}%
  \BibitemOpen
  \bibfield  {author} {\bibinfo {author} {\bibfnamefont {E.~D.}\ \bibnamefont
  {Palik}},\ }\href@noop {} {\  (\bibinfo {year} {1998})}\BibitemShut {NoStop}%
\bibitem [{\citenamefont {Gelfand}\ and\ \citenamefont
  {Shilov}(2017)}]{gelfand2}%
  \BibitemOpen
  \bibfield  {author} {\bibinfo {author} {\bibfnamefont {I.~M.}\ \bibnamefont
  {Gelfand}}\ and\ \bibinfo {author} {\bibnamefont {Shilov}},\ }\href@noop {}
  {\bibfield  {journal} {\bibinfo  {journal} {Providence, Rhode Island : AMS
  Chelsea Publishing}\ } (\bibinfo {year} {2017})}\BibitemShut {NoStop}%
\bibitem [{\citenamefont {Arbabi}\ \emph {et~al.}(2015)\citenamefont {Arbabi},
  \citenamefont {Horie},\ and\ \citenamefont {Faraon}}]{art22icle}%
  \BibitemOpen
  \bibfield  {author} {\bibinfo {author} {\bibfnamefont {A.}~\bibnamefont
  {Arbabi}}, \bibinfo {author} {\bibfnamefont {Y.}~\bibnamefont {Horie}}, \
  and\ \bibinfo {author} {\bibfnamefont {A.}~\bibnamefont {Faraon}},\
  }\href@noop {} {\bibfield  {journal} {\bibinfo  {journal} {Nature
  nanotechnology}\ }\textbf {\bibinfo {volume} {10}} (\bibinfo {year}
  {2015})}\BibitemShut {NoStop}%
\bibitem [{\citenamefont {Zangeneh-Nejad}\ and\ \citenamefont
  {Khavasi}(2017)}]{zangeneh2017spatial}%
  \BibitemOpen
  \bibfield  {author} {\bibinfo {author} {\bibfnamefont {F.}~\bibnamefont
  {Zangeneh-Nejad}}\ and\ \bibinfo {author} {\bibfnamefont {A.}~\bibnamefont
  {Khavasi}},\ }\href@noop {} {\bibfield  {journal} {\bibinfo  {journal}
  {Optics letters}\ }\textbf {\bibinfo {volume} {42}},\ \bibinfo {pages} {1954}
  (\bibinfo {year} {2017})}\BibitemShut {NoStop}%
\end{thebibliography}%






\end{document}